\documentclass{aa}  

\usepackage{graphicx}
\usepackage{txfonts}
\usepackage{natbib}
\usepackage[draft]{hyperref}
\hypersetup{
    colorlinks=true,
    linkcolor=blue,
    filecolor=magenta,      
    urlcolor=blue,
    citecolor=blue,
}
\begin{document}

   \title{Formation of moon systems around giant planets}

   \subtitle{Capture and ablation of planetesimals as foundation for a \\ pebble accretion scenario}

   \author{T. Ronnet
          \and
          A. Johansen
          }

   \institute{Department of Astronomy and Theoretical Physics, Lund Observatory, Lund University, Box 43, 22100 Lund, Sweden\\
              \email{thomas.ronnet@astro.lu.se}
             }

   \date{Received ; accepted }
 
  \abstract{The four major satellites of Jupiter, known as the Galilean moons, and Saturn's most massive satellite, Titan, are believed to have formed in a predominantly gaseous circum-planetary disk, during the last stages of formation of their parent planet. Pebbles from the protoplanetary disk are blocked from flowing into the circumplanetary disk by the positive pressure gradient at the outer edge of the planetary gap, so the gas drag assisted capture of planetesimals should be the main contributor to the delivery of solids onto circum-planetary disks. However, a consistent framework for the subsequent accretion of the moons remains to be built. Here we use numerical integrations to show that most planetesimals being captured within a circum-planetary disk are strongly ablated due to the frictional heating they experience, thus supplying the disk with small dust grains, whereas only a small fraction 'survives' their capture. We then construct a simple model of a circum-planetary disk supplied by ablation, where the flux of solids through the disk is at equilibrium with the ablation supply rate, and investigate the formation of moons in such disks. We show that the growth of satellites is driven mainly by accretion of the pebbles that coagulate from the ablated material. The pebble-accreting protosatellites rapidly migrate inward and pile up in resonant chains at the inner edge of the circum-planetary disk. We propose that dynamical instabilities in these resonant chains are at the origin of the different architectures of Jupiter's and Saturn's moon systems. The assembly of moon systems through pebble accretion can therefore be seen as a down-scaled manifestation of the same process that forms systems of super-Earths and terrestrial-mass planets around solar-type stars and M-dwarfs.}


   \maketitle
%

\section{Introduction}

The formation of massive moons around gas giant planets is envisioned to take place in a gaseous disk surrounding the planet in the last stages of its accretion \citep[see, e.g.,][for a review]{PC15}.
Their formation would thus be analogous to that of planets around stars and face the same theoretical challenges. 
These include for example the formation of satellitesimals to seed the accretion of more massive moons and the survival of these latter against their rapid inward migration.
In addition, some issues, such as the delivery of solids to the circum-planetary disk (hereafter CPD), are specific to the formation of satellites around the giant planets.
The observed properties of the Galilean moons orbiting Jupiter, as well as that of Saturn's moons, provide additional constraints on the conditions under which they have formed.

The satellite systems of Jupiter and Saturn both represent a similar fraction of the mass of their parent planets ($\sim$10$^{-4}\,M_\mathrm{p}$) and are quite compact \citep[e.g.,][]{CW06}.
In the case of the Galilean system, the three inner moons--Io, Europa and Ganymede--form a resonant chain (known as a Laplace resonant system).
Additionally, the decreasing bulk densities of the Galilean satellites with respect to their distance from Jupiter is suggestive of a compositional gradient among the moons, with increasingly volatile rich compositions away from Jupiter \citep[see, e.g.,][]{HSL15}.
Finally, information regarding the internal structure of the moons has been inferred from the gravity measurements performed by the \textit{Galileo} spacecraft at the jovian system and the \textit{Cassini} spacecraft in the case of Saturn's moons.
While the three inner Galilean moons (involved in the Laplace resonance system) are likely to be fully differentiated with the presence of an iron core, a silicate mantle and, in the case of Europa and Ganymede, an icy outer mantle, Callisto and Titan appear to be only partially differentiated \citep{An+01,Iess+10}.
The formation of undifferentiated satellites implies limited heating during accretion to prevent large scale ice melting and hence requires long formation timescales \citep[$\gtrsim$1$\,$Myr;][]{BC08}.
Even so, a later differentiation of the satellites could be difficult to avoid if compositional gradients prevent an efficient transport of radiogenic heating through convection \citep{OS14}. 
Also, non-hydrostatic effects could have an important impact on the derivation of the moment of inertia of slowly rotating satellites such as Callisto and Titan \citep{GS13}, so that the interpretation of their internal structure remains uncertain and these objects might in fact be differentiated.
If this is the case, these satellites could have formed over much shorter timescales.

The consideration of the properties of the satellite systems, together with the available knowledge on processes relevant to planet formation, guided the development and refinement of satellite formation scenarios \citep[e.g.,][]{LS82,CFFM95,CW02,CW06,ME03a,ME03b,MGB02,SSI10}.
Although the gas-starved model developed by \citet{CW02,CW06} has been well recognized as a plausible satellite formation scenario, in the recent years, new paradigms have emerged regarding several key processes for the formation of planets which challenge our current understanding of the formation of the giant planets' moons.

These challenges are briefly discussed in Section~\ref{Motiv}, where we argue that the capture and ablation of planetesimals should be an important source of solids in giant planet's CPD, as previously suggested \citep[e.g.,][]{Es+09}, but unlike the assumption of the gas-starved model.
Such a mechanism is investigated in Section~\ref{Capt_abl}, which shows that most of the planetesimals captured in the CPD are efficiently ablated.
Ultimately, the material ablated off of the surface of the planetesimals provides a source  of dust in the CPD whose subsequent evolution is investigated in Section~\ref{Dust_evol}.
These results provide the ground for the development of a revised formation model for the giant planets' satellites (Section~\ref{accretion}). Specifically, we propose that the seeds of the satellites initially form from the fraction of captured planetesimals that survived ablation in the CPD and subsequently grow through pebble accretion from the flux of dust supplied by the  ablation of planetesimals.
More aspects and implications are discussed in Section~\ref{discussion}, before we summarize our results in Section~\ref{sum}.

\section{Motivation}\label{Motiv}

The so-called gas-starved model developed by \citet{CW02,CW06} has been the leading scenario for the formation of satellites around gas giant planets because of its apparent ability to resolve long-standing issues which are primarily i) the impossibility of satisfying the minimum mass requirement to grow the moons and the low temperature needed for ice stability simultaneously in a 'self-consistent' viscous disk model \citep{MDR99,CW02}, ii) the difficulty of forming undifferentiated massive satellites and iii) the survival of the satellites against their inward migration.

In this scenario, the formation of the satellites takes place as the giant planet is still accreting material from the surrounding protoplanetary disk (PPD).
The CPD is not a closed system then, but is instead constantly replenished by fresh material with an approximately solar composition (dust-to-gas mass ratio $\varepsilon_{\mathrm{d},\odot}\approx 10^{-2}$).
In this case, all the mass required to grow the satellites need not be present at one time in the CPD, which allows to consider less dense environments which are more propitious to cold temperature and the formation of icy satellites.
The accretion timescale of the moons would be regulated by the rate of inflow of fresh material onto the CPD, and the migration timescales would be lengthen due to the lower gas densities. 

However, it should be noted that the validity of viscous disk models has been challenged in the recent years \citep[see, e.g.,][]{Tu+14,GTNM15,Bai17} due to the fact that non-ideal MHD effects tend to suppress the source of turbulent viscosity in the disk, and their evolution would then rather be driven by thermo-magnetic winds.
This possibly results in inefficient viscous heating of the disks \citep{MBO19}.
The non-ideal MHD effects are expected to dominate even more in the case of CPDs \citep{FOTI14,FKTG17}, so that CPDs could be denser than advocated by \citet{CW02} while remaining cold.
Moreover, the truncation of the giant planets' CPD by an inner magnetic cavity, as seems to be required to explain their rotation rate \citep{TS96,Ba18}, would stop the migration of the satellites \citep{SSI10,OI12}.
It thus seems that a gas-starved environment is not essential to allow for the formation of icy satellites and their survival.

On the other hand, one of the important issues of the gas-starved model is the formation of satellitesimals and satellites seeds.
Whereas it has been customarily assumed that large objects would form out of the small dust grains in the CPD \citep[e.g.,][]{CW02,CW06,SSI10,OI12}, it is now understood that the formation of planetesimals or satellitesimals likely requires some instability (e.g., streaming instabilities) or adequate environment (e.g., pressure bumps) to allow for the efficient concentration of dust that can then gravitationally collapse into 10--100$\,$km sized objects \citep[see][for a review]{Jo+14}.
The rapid inward drift timescale of dust grains in a CPD precludes the formation of satellitesimals through any known mechanism \citep{SOSI17}.
Even if satellitesimals were somehow able to form \citep[see, for example,][]{DS18}\footnote{These authors proposed that satellitesimal formation in CPDs was possible thanks to the existence of a dust trap arising from the complex radial flow of gas observed in 3D viscous simulations. However, at low viscosities, this flow pattern should vanish \citep{Sz+14}, and, whereas the flow pattern depends on the height above the midplane and reverses, the evolution of the dust was restricted to the midplane of the disk. Thus, more investigations would be needed to assess the robustness of the dust trap.}, it is not expected that all the dust grains would be converted into large objects, contrary to the assumption in previous investigations. 
Rather, the satellitesimals should further grow by accreting the remaining (and possibly re-supplied) dust grains in a process known as pebble accretion \citep[see the recent reviews by][]{JL17,Or17}.
The accretion of moons in CPDs thus needs to be reassessed considering this more likely growth channel.

Another important issue is that it is unlikely that the material accreted by a giant planet in the late stages of its formation has a solar dust-to-gas mass ratio, as advocated in the gas-starved models \citep[see][for a discussion]{Ro18}.
Above a mass of typically a few $\sim$10$\,M_\oplus$, a protoplanet starts to significantly perturb the gas disk in its vicinity, opening a shallow gap whose outer edge acts as a barrier for drifting dust grains \citep{MN12,LJM14}.
At a mass comparable to that of Jupiter, dust grains larger than a few 10--100$\,\mu$m should be efficiently filtered \citep{Paa07,ZNDEH12,Bi+18,WBGKP18}, likely resulting in a drop of the dust-to-gas mass ratio of the accreted material at the order of magnitude level.
Combined with the fact that only a small fraction of the dust grains would be accreted by the forming moons (see Sect.~\ref{peb_eff} for an estimate), it becomes difficult to envision that the inflow of small dust grains entrained with the gas constitutes the main source of material for the formation of moons.

\citet{Es+09} proposed that gas-drag assisted capture and ablation of planetesimals on initially heliocentric orbits in the vicinity of the giant planets could have provided the solids necessary to form the moons.
This mechanism has recently received some attention \citep{MEC10,FOTS13,TMM14,DAP15,SO17,Ro18}, although it has not been put in the perspective of constructing a more consistent scenario of the subsequent formation of satellites.
Here we therefore investigate the capture and ablation of planetesimals in the CPD of a jovian mass planet with the aim to parametrize the distribution of material (Sect.~\ref{Capt_abl}) so as it could be used as an input for models of the subsequent accretion of satellites (Sect.~\ref{Dust_evol} and \ref{accretion}).
We argue that the efficient ablation of planetesimals provides the ground for a pebble accretion scenario for the formation of giant planets' moons, in many regards similar to the scenario proposed for the formation of compact super-Earth systems \citep{OLS17, Iz+17, Iz+19, La+19}.

 We note that a pebble accretion scenario for the origin of the Galilean satellites, which shares many similarities with the ideas presented in Sect.~\ref{accretion}, has recently been investigated by \citet{Shib+19}. However, similarly to \citet{CW02}, these authors assumed that dust grains were brought to the CPD with the gas accreted by Jupiter, which is unlikely to provide a sufficient source of solid material as discussed above. Moreover, the coupling between the accretion rate of gas and dust onto the CPD implies that the Galilean moons must have started to form as Jupiter was only 40\% of its present mass, and that Jupiter accreted over a very long timescale of 30 Myr--whereas the typical lifetime of protoplanetary disks is only $\sim$3 Myr \citep{HHC16}, to reproduce their current properties. If on the other hand the building blocks of the moons are brought to the CPD through the capture and ablation of planetesimals, as proposed here, the formation of the satellites can take place much later in the accretion history of their parent planet, and these latter are not constrained to form on several tens of Myr.

\section{Circum-planetary disk model}\label{CPD_model}

Based on the recent findings that the turbulence level inside a gaseous CPD is expected to be low \citep{FOTI14,FKTG17}, and that the midplane of such disks are inefficiently heated by internal dissipation mechanisms \citep{MBO19}, we here consider the case of a passively irradiated CPD.

We assume that the surface density of the disk follows a power-law such that
\begin{equation}\label{Sigma_CPD}
	\Sigma_\mathrm{g} = \Sigma_\mathrm{out} \left(\frac{r}{r_\mathrm{out}} \right)^{-\gamma},
\end{equation}
where $r$ is the radial distance to the planet in cylindrical coordinates, $r_\mathrm{out}$ is the outer radius of the CPD and
\begin{equation}
	\Sigma_\mathrm{out} = \frac{(2-\gamma)M_\mathrm{CPD}}{2\pi r^2_\mathrm{out}}
\end{equation}
is the surface density of gas at the outer edge of the disk and $M_\mathrm{CPD}$ is the total mass of the CPD. 
Here we adopt $r_\mathrm{out}=0.2\,R_\mathrm{H}$ (with $R_\mathrm{H}$ the size of the Hill sphere of the planet) as reported by \citet{TOM12} from the analysis of their 3D hydrodynamic simulations. 
Although we note that there is no sharp boundary at the outer edge of the CPD allowing to precisely define its radial extent \citep[e.g.,][]{TOM12,Sz+14}, both the capture/ablation of planetesimals and the subsequent growth of satellites take place in the inner part of the CPD in our scenario (e.g., mainly interior to $r \lesssim 0.05 R_\mathrm{H}$).
We assume that the mass of the CPD represents some fraction of the mass of the planet, $M_\mathrm{CPD}=1.5\times10^{-3}\,M_\mathrm{p}$, in agreement with recent hydrodynamic simulations \citep{Sz17}.
We set the index of the power-law of the gas surface density at $\gamma=1.5$, which is also in agreement with the simulation results of \citet{TOM12} and \citet{Sz17}\footnote{as actually reported in \citet{DS18}}.

The midplane temperature $T_\mathrm{d}$ of a passively irradiated disk depends on the angle at which the light from the central radiating object strikes the disk, or grazing angle.
It can be approximated by \citep{RP91}
\begin{equation}\label{T_disk}
	T_\mathrm{d} = T_\mathrm{p} \left[\frac{2}{3\pi}\left(\frac{R_\mathrm{p}}{r} \right)^3 + \frac{1}{2} \left(\frac{R_\mathrm{p}}{r} \right)^2 \frac{H_\mathrm{ph}}{r}\left(\frac{d \ln H_\mathrm{ph}}{d \ln r}-1 \right) \right]^{1/4},
\end{equation}
 where $T_\mathrm{p}$ and $R_\mathrm{p}$ are the temperature and radius of the central planet, respectively, and $H_\mathrm{ph}$ is the photospheric height of the disk at which radiation from the planet is intercepted.  
 This latter quantity is usually assumed to be proportional to the pressure scale height of the disk, $H_\mathrm{ph}=\chi H_\mathrm{g}$, where $H_\mathrm{g}=c_\mathrm{g}/\Omega_\mathrm{K}$ (with $c_\mathrm{g}$ the isothermal speed of sound in the gas and $\Omega_\mathrm{K}$ the orbital frequency), and the proportionality factor $\chi$, which is of order unity, and depends on the opacity of the disk \citep[e.g.,][]{CG97,DDN01}.
 Assuming that gas accretion onto the giant planet is still ongoing, $T_\mathrm{p}$ is related to the mass accretion rate onto the planet, since the planet's luminosity, $L_\mathrm{p}=4\pi R^2_\mathrm{pl}\sigma_\mathrm{sb}T^4_\mathrm{p}$, is then given by the accretion luminosity $L_\mathrm{p} \approx GM_\mathrm{p}\dot{M}_\mathrm{p}/R_\mathrm{p}$ \citep[e.g.,][]{MMM17}.
 To remain as general as possible, it is convenient to parameterize the accretion rate onto the planet through the accretion timescale $\tau_\mathrm{acc} \equiv M_\mathrm{p}/\dot{M}_\mathrm{p}$.
 The planet's radiating temperature is then $T_\mathrm{p}\propto M^{1/2}_\mathrm{p}R^{-3/4}_\mathrm{p}\tau^{-1/4}_\mathrm{acc}$.

 We can now use eq.~\ref{T_disk} to derive some scaling of the temperature of the CPD as a function of the distance from the planet. 
 The first term inside the brackets of eq.~\ref{T_disk} accounts for the finite angular size of the radiating planet, whereas the second term arises from the flaring geometry of the disk.
 In the inner portions of the CPD, where $H_\mathrm{ph} \lesssim R_\mathrm{p}$, the first term will dominate, while at larger distances, the grazing angle is instead determined by the flaring of the disk.
 There are thus two asymptotic solutions for the temperature of the CPD.
 In the inner, geometrically thin (with respect to the planet's size), portion of the CPD, the temperature follows
 \begin{equation}
 T_\mathrm{thin} \approx 190 \left(\frac{M_\mathrm{p}}{M_\mathrm{Jup}} \right)^{1/2} \left(\frac{\tau_\mathrm{acc}}{5\,\mathrm{Myr}} \right)^{-1/4} \left(\frac{r}{10\, R_\mathrm{Jup}} \right)^{-3/4}\mathrm{K},
 \end{equation}
which has the same dependency on distance ($T\propto r^{-3/4}$) than viscous disk models \citep[e.g.,][]{RP91,CW02}.
In the regions where flaring dominates instead, the temperature becomes
\begin{equation}
\begin{split}
	T_\mathrm{flar} \approx & 190 \left(\frac{r}{10\,R_\mathrm{Jup}} \right)^{-3/7} \\
	{} & \times \left(\frac{M_\mathrm{p}}{M_\mathrm{Jup}} \right)^{3/7}  \left(\frac{R_\mathrm{p}}{1.6\,R_\mathrm{Jup}} \right)^{-2/7} \left(\frac{\tau_\mathrm{acc}}{5\,\mathrm{Myr}} \right)^{-2/7} \left( \frac{\chi}{4}\right)^{2/7}\mathrm{K}, 
\end{split}
\end{equation}  
which follows the classical result derived by \citet{CG97} that $T\propto r^{-3/7}$.
The temperature of the CPD can then be approximated by $T_\mathrm{d}=\max(T_\mathrm{thin},T_\mathrm{flar})$. 
By equating the two asymptotic solutions, we find that the transition between the two regimes should occur at a distance
\begin{equation}
r_\mathrm{tran} \approx 10 \left(\frac{M_\mathrm{p}}{M_\mathrm{Jup}} \right)^{2/9} \left(\frac{\tau_\mathrm{acc}}{5\,\mathrm{Myr}} \right)^{1/9} \left(\frac{R_\mathrm{p}}{1.6\, R_\mathrm{Jup}} \right)^{8/9} \left(\frac{\chi}{4} \right)^{-8/9} R_\mathrm{Jup}.
\end{equation}
The resulting aspect ratio of the CPD, $h_\mathrm{g}\equiv H_\mathrm{g}/r$, is
\begin{equation}\label{h_thin}
h_\mathrm{thin} \approx 0.06 \left(\frac{M_\mathrm{p}}{M_\mathrm{Jup}} \right)^{-1/4} \left(\frac{\tau_\mathrm{acc}}{5\, \mathrm{Myr}} \right)^{-1/8} \left( \frac{r}{10\, R_\mathrm{Jup}}\right)^{1/8}, 
\end{equation}
when $r \leqslant r_\mathrm{tran}$, and 
\begin{equation}
\begin{split}
h_\mathrm{flar} \approx & 0.06 \left( \frac{r}{10\, R_\mathrm{Jup}}\right)^{2/7}  \\
{} & \times \left(\frac{M_\mathrm{p}}{M_\mathrm{Jup}} \right)^{-2/7} \left(\frac{\tau_\mathrm{acc}}{5\, \mathrm{Myr}} \right)^{-1/7} \left(\frac{R_\mathrm{p}}{1.6\, R_\mathrm{Jup}} \right)^{-1/7} \left(\frac{\chi}{4} \right)^{1/7},
\end{split}
\end{equation}
for $r > r_\mathrm{tran}$.

In the above estimates, we have used a rather long accretion timescale, which is similar to the value adopted by \citet{CW02,CW06} at the formation epoch of the Galilean moons, and considered a slightly inflated planet, as expected at the end of the runaway contraction phase \citep[e.g.,][]{LHAB09,MMM17}.
The resulting luminosity, for a jovian mass planet, would be $L_\mathrm{p}\approx 4\times10^{-5}\, L_\odot$, which is also consistent with numerical simulations \citep{LHAB09,MMM17}. 
For simplicity, we have here adopted a constant photospheric to pressure scale height ratio $\chi =4$, corresponding to the typical value reported by \citet{CG97} for a disk with a solar metallicity. 
However, we elaborate in Section~\ref{Dust_evol} on the fact that the dust-to-gas mass ratio in a CPD is likely to be much lower than expected from a solar composition mixture.
A further improvement would thus be to self-consistently determine $\chi$ \citep[as in][]{DDN01} based on the dust spatial and size distribution, but we leave such considerations to future investigations.

Applying this passive CPD model to the case of Saturn, we can note some differences in the resulting thermal structure in the region of satellites formation.
The temperature scales to the Saturn system as
\begin{equation}
	T_\mathrm{thin} \approx 120 \left(\frac{M_\mathrm{p}}{M_\mathrm{Sat}} \right)^{1/2} \left(\frac{\tau_\mathrm{acc}}{5\,\mathrm{Myr}} \right)^{-1/4} \left(\frac{r}{10\, R_\mathrm{Sat}} \right)^{-3/4}\mathrm{K},
\end{equation}

\begin{equation}
\begin{split}
	T_\mathrm{flar} \approx &  130\, \mathrm{K} \, \left(\frac{r}{10\,R_\mathrm{Sat}} \right)^{-3/7} \left(\frac{M_\mathrm{p}}{M_\mathrm{Sat}} \right)^{3/7} \\
	{} & \times  \left(\frac{R_\mathrm{p}}{1.6\,R_\mathrm{Sat}} \right)^{-2/7} \left(\frac{\tau_\mathrm{acc}}{5\,\mathrm{Myr}} \right)^{-2/7} \left( \frac{\chi}{4}\right)^{2/7}, 
\end{split}
\end{equation}
which is overall colder than in the case of the jovian disk.
The resulting aspect ratio of the CPD is on the contrary larger, due to the lower planet's mass, where
\begin{equation}
h_\mathrm{thin} \approx 0.079 \left(\frac{M_\mathrm{p}}{M_\mathrm{Sat}} \right)^{-1/4} \left(\frac{\tau_\mathrm{acc}}{5\, \mathrm{Myr}} \right)^{-1/8} \left( \frac{r}{10\, R_\mathrm{Sat}}\right)^{1/8}, 
\end{equation}
\begin{equation}
\begin{split}
h_\mathrm{flar} \approx \, & 0.082 \left( \frac{r}{10\, R_\mathrm{Sat}}\right)^{2/7} \\
{} & \times \left(\frac{M_\mathrm{p}}{M_\mathrm{Sat}} \right)^{-2/7} \left(\frac{\tau_\mathrm{acc}}{5\, \mathrm{Myr}} \right)^{-1/7} \left(\frac{R_\mathrm{p}}{1.6\, R_\mathrm{Sat}} \right)^{-1/7} \left(\frac{\chi}{4} \right)^{1/7}.
\end{split}
\end{equation}
The transition between the two solutions occurs at a distance
\begin{equation}
r_\mathrm{tran} \approx 7.8 \left(\frac{M_\mathrm{p}}{M_\mathrm{Sat}} \right)^{2/9} \left(\frac{\tau_\mathrm{acc}}{5\,\mathrm{Myr}} \right)^{1/9} \left(\frac{R_\mathrm{p}}{1.6\, R_\mathrm{Sat}} \right)^{8/9} \left(\frac{\chi}{4} \right)^{-8/9} R_\mathrm{Sat}.
\end{equation}
   
\section{Capture and ablation of planetesimals}\label{Capt_abl}

  \subsection{Methods}
  
\begin{figure}
  \centering
  \includegraphics[width=\linewidth]{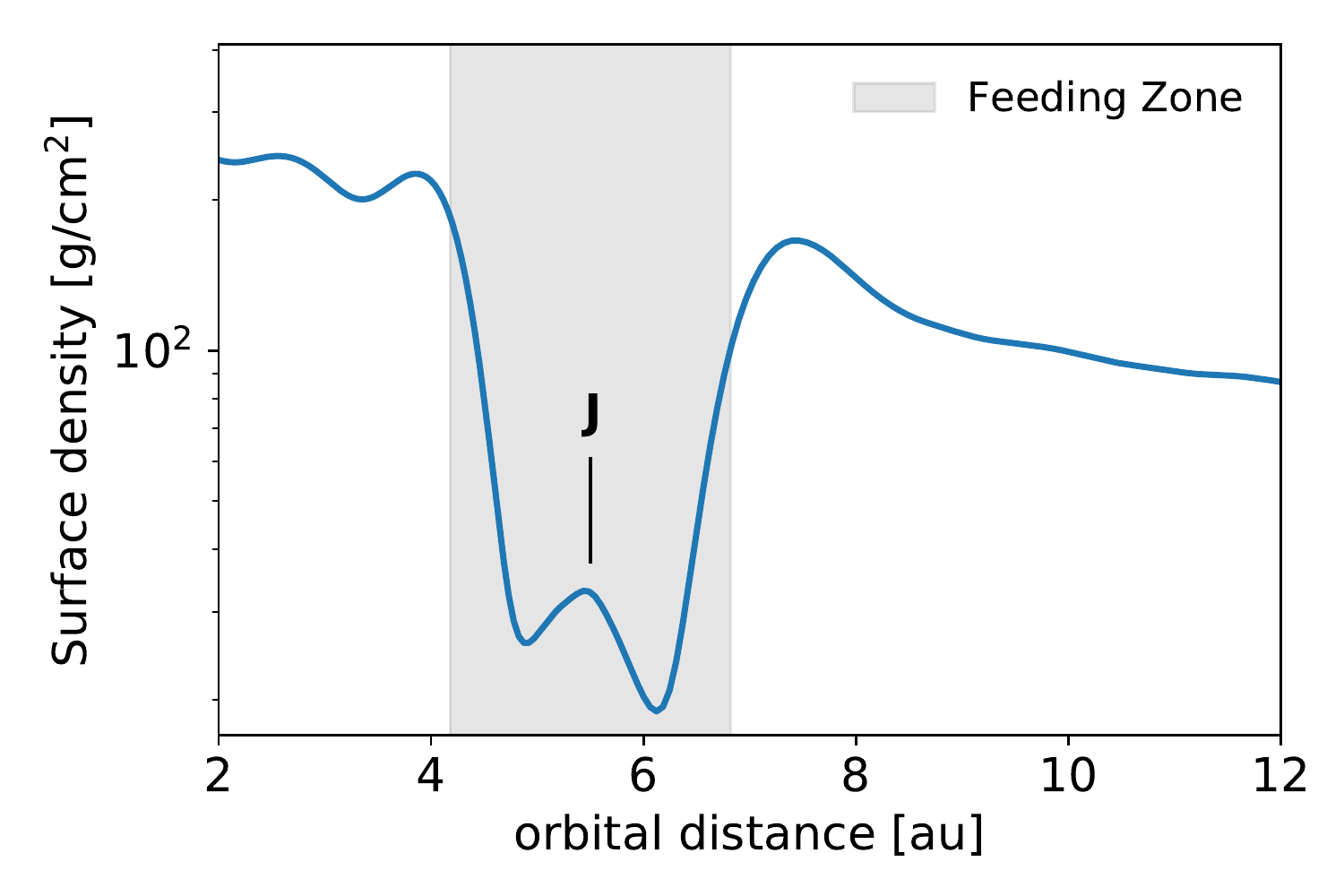}
  \caption{Surface density profile of the protoplanetary disk obtained by a 2-dimensional hydrodynamic simulation of a Jupiter mass planet in a disk with a constant aspect ratio $H_\mathrm{PPD}/R = 0.05$.}\label{SD_PPD}
  \end{figure}

  We model the evolution of a swarm of planetesimals embedded in a protoplanetary disk and orbiting close to a Jupiter-mass planet located at 5.5 au from a Sun-mass star. 
Fig.~\ref{SD_PPD} shows the surface density of gas in the vicinity of the planet (see below for details) along with the region where planetesimals were initially distributed, which is shown as a grey shaded area and corresponds to the so-called feeding zone of the planet.    
 Following \citet{SI08}, we define the feeding zone as the region where the Jacobi energy of the planetesimals $E_\mathrm{J}$ is positive. $E_\mathrm{J}$ can be expressed as \citep[e.g.,][]{NIN89,SI08}
  \begin{equation}
  E_\mathrm{J}=\frac{1}{2}\left[\left(\frac{e}{h}\right)^2+\left(\frac{i}{h}\right)^2\right] - \frac{3}{8}b^2 + \frac{9}{2} + O(h).
  \end{equation}
  In the above expression, $h=R_\mathrm{H}/a_p=(M_p/3M_\odot)^{1/3}$ is the reduced Hill radius of the planet, with $M_p$ its mass and $a_p$ its orbital distance, $e$ and $i$ are the eccentricity and inclination of the planetesimal, and $b$ is the non-dimensional orbital separation with the planet defined as
  \begin{equation}
  b=\frac{a-a_p}{h a_p},
  \end{equation}
  where $a$ is the (heliocentric) semimajor axis of the planetesimal.
  When the eccentricity and inclination distributions of the planetesimals are small (i.e., both $e/h$ and $i/h\lesssim 1$), the feeding zone of the planet extends to $b\approx \pm 2\sqrt{3}$. 
 In our simulations, 3000 planetesimals are uniformly randomly distributed in the feeding zone of the planet with initial eccentricities and inclinations drawn from a Rayleigh distribution with a mean of $10^{-3}$ and $5\times10^{-4}$, respectively.
The planetesimals were considered as massless test particles and the orbits of all the bodies were integrated using the adaptive time stepping \texttt{IAS15} integrator provided within the \texttt{REBOUND}\footnote{\url{ http://github.com/hannorein/rebound}} package \citep{RS15}.
Planetesimals were assigned a physical size to account for the effect of gas drag on their orbital evolution (see next paragraph), and we performed runs for planetesimals with sizes of either 100, 10 or $1\,$km.
The implementation of the effect of gas drag and ablation of planetesimals is described in the following paragraphs.

\paragraph{Planetesimal dynamics} Since we are here interested in the gas drag assisted capture of planetesimals, the effect of gas friction on the evolution of the objects was accounted for using the stopping time formalism \citep[e.g.,][]{Wh72}, where the following acceleration was added to the equation of motion of the planetesimals,
\begin{equation}\label{a_drag}
	\mathbf{a}_\mathrm{drag}=-\frac{\mathbf{v}-\mathbf{v}_g}{t_s}.
\end{equation} 
Here, $\mathbf{v}$ is the velocity vector of the planetesimal and $\mathbf{v}_g$ that of the gas. The stopping time $t_s$ is defined by \citep[e.g.,][]{PMC11,GIO14}
\begin{equation}\label{tstop}
	t_s = \left(\frac{\rho_g v_\mathrm{th}}{\rho_\mathrm{pl}R_\mathrm{pl}} \min\left[ 1, \frac{3}{8}\frac{v_\mathrm{rel}}{v_\mathrm{th}}C_D\right]  \right)^{-1},
\end{equation}
where $R_\mathrm{pl}$ and $\rho_\mathrm{pl}$($=1000\,\mathrm{kg\,m}^{-3}$) are the radius and internal density of the planetesimal, respectively, $v_\mathrm{rel}$ is the magnitude of the relative velocity between the planetesimal and the gas, $\rho_g$ is the density of the gas disk (either the PPD or the CPD),  $v_\mathrm{th} = \sqrt{8/\pi}c_g$ is the thermal velocity of the gas, and $C_D$ is the dimensionless drag coefficient.
Although $C_D$ is in principle a complicated function of the Reynolds and Mach numbers of the flow around an object \citep[for example, see][]{DAP15}, it is of order unity at large Mach and Reynolds numbers \citep{TMM14,SO17}. Similarly to \citet{SO17}, we therefore opted to set $C_D=1$.

\paragraph{Planetesimal thermodynamics} To estimate the effect of the ablation of planetesimals on the delivery of solid material to the CPD, the evolution of the surface temperature of the objects must be followed. 
To do so, we assume that the surface temperature $T_\mathrm{pl}$ of the planetesimals is always at equilibrium with respect to heating from radiation at the ambient temperature of the CPD and  friction with the gas, and cooling due to the release of the latent heat of vaporization of water (see Appendix~\ref{App} for details). This yields the following expression for the temperature
\begin{equation}\label{Tsurf}
	T^4_\mathrm{pl} = T^4_\mathrm{d}+ \frac{C_D \rho_g v^3_\mathrm{rel}}{32\sigma_\mathrm{sb}} -\frac{P_v(T_\mathrm{pl})}{\sigma_\mathrm{sb}} \sqrt{ \frac{\mu_\mathrm{H_2O} }{ 8 \pi R_g T_\mathrm{pl}}}L_w,
\end{equation}
where $T_\mathrm{d}$ is the local temperature of the CPD (assumed to be isothermal in the vertical direction), $\sigma_\mathrm{sb}$ is the Stefan-Boltzmann constant, $R_g$ is the ideal gas constant, $\mu_\mathrm{H_2O}=0.018\,\mathrm{kg\,mol}^{-1}$ is the molecular weight of water, $L_w=3\times10^{6}\, \mathrm{J\,kg}^{-1}$ is the latent heat of vaporization of water and $P_v$ is the saturated vapor pressure of water and is expressed as polynomials of the temperature \citep[taken from][]{FS09}.
Here we focus on the ablation of water only, assuming that small silicate grains would be entrained with the water released from the surface of the planetesimals.
This is only accurate if the planetesimals are not differentiated, and water ice and silicates are well mixed. 
If the planetesimals are differentiated into an icy mantle and rocky core, then the ablation of the silicate core should be considered \citep{MEC10}.
We show in Fig.~\ref{abl_ice_sil} that such a consideration should not have a great influence on the results presented here (see figure caption for details).

The cooling of the surface due to the ablation of water ice (third term on the r.h.s of eq.~\ref{Tsurf}) depends itself upon the surface temperature of the object.
We must therefore solve eq.~\ref{Tsurf} iteratively to determine the surface temperature of planetesimals. When $T_\mathrm{pl}$ is known, the mass ablation rate from the surface of the object can be expressed as \citep[e.g.,][]{DAP15},
\begin{equation}
	\dot{m}_\mathrm{abl}=-4\pi R^2_\mathrm{pl} P_v(T_\mathrm{pl}) \sqrt{\frac{\mu_\mathrm{H_2O}}{2\pi R_g T_\mathrm{pl}}}.
\end{equation}
The ablation of material also changes the size of planetesimals, which affects their dynamics through a change in their stopping time (eq.~\ref{tstop}). 

We solve eq.~\ref{Tsurf} in between each timestep of the N-body integrator and evaluate the mass loss and radius change due to ablation for each planetesimal that interacts with the CPD (we ignore ablation of material in the PPD). Our approach is not fully numerically self-consistent in the sense that the timestep of the integrator changes according to the dynamical evolution of the objects only, regardless how much mass may be lost (equivalently, size may be reduced) during the timestep. However, as we show in the following, planetesimals are mostly ablated during their closest approach to the planet or, in the case of captured objects, around the pericenter of their orbit. This is also when the integrator timestep is the smallest, of the order of $10^{-4}$--$10^{-5}\, \mathrm{yr/2\pi}$. Most strongly heated objects can reach surface temperatures of $T_\mathrm{pl}\sim550\,$K, which, for a planetesimal with a size $R_\mathrm{pl}=100\,$km, would imply a mass loss of the order of $10^{-2}$--$10^{-3}$ of its total mass during a typical timestep.
  
  \begin{figure*}
	\includegraphics[width=\linewidth]{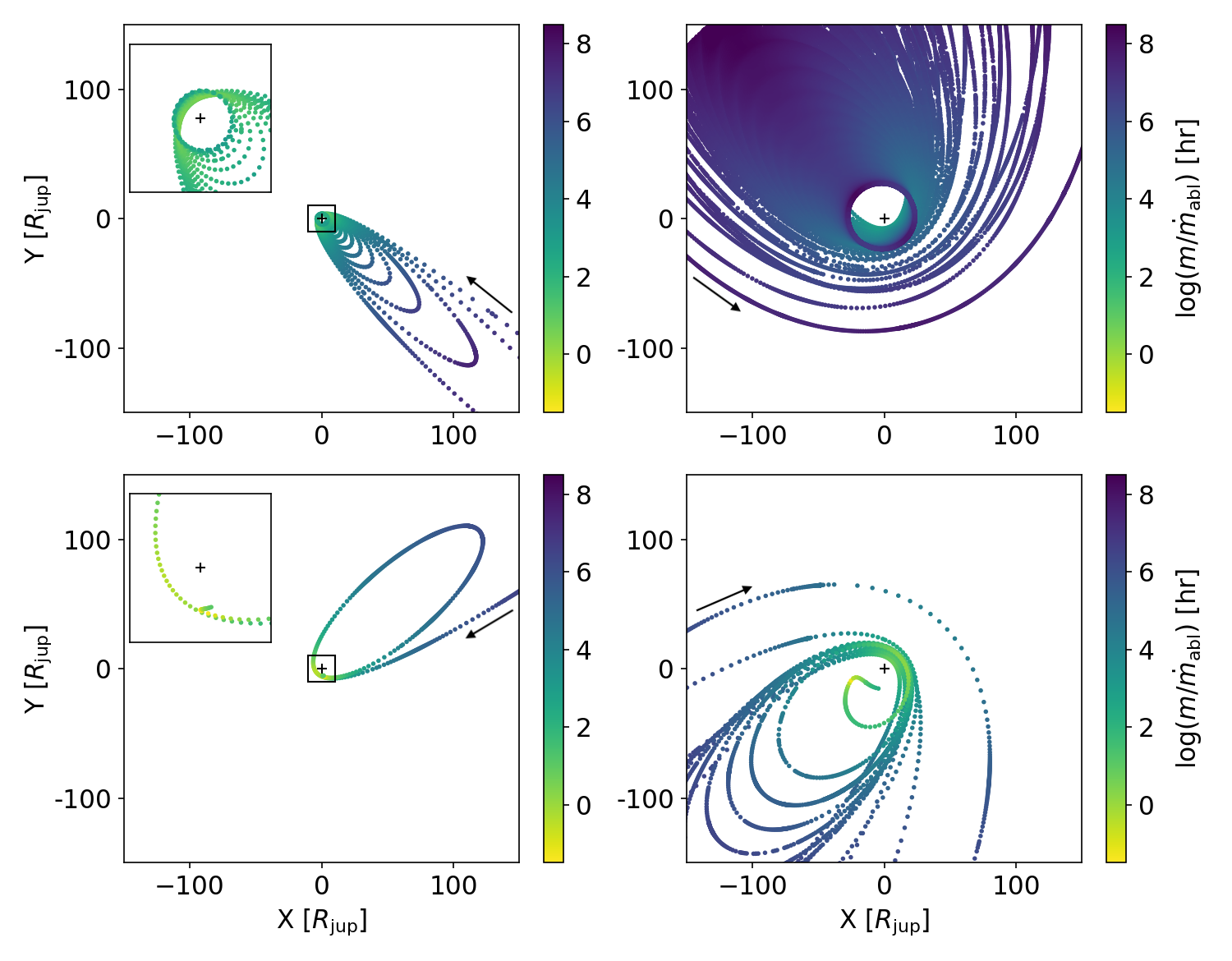}
	\caption{Example of orbits of objects that are captured in the prograde (top panels) and retrograde (bottom panels) directions in the planetocentric frame. Each dot corresponds to the position of the planetesimal at a given timestep and the arrows mark the direction at which the objects approach the planet. The left row shows the 'typical' orbital evolution during the gas drag assisted capture of planetesimals whereas the right row corresponds to the orbital evolution following lower probability events where planetesimals are captured on orbits with a large pericenter and their semimajor axis decays slowly \citep[see Fig.~\ref{a_Rpl}; see also,][]{SO17}. Insets on the top and bottom left panels are close-up views of the trajectories near the pericenter of the orbits with $\mathrm{X,Y}\in[-10;10]$. The color of the dots indicates the ablation rate from the surface of the planetesimal. Ablation occurs mainly around the point of closest approach to the planets where the density of gas is the highest. Due to their larger relative velocities with respect to the gas, and hence larger frictional heating, planetesimals on retrograde orbits (bottom) experience more sustained ablation with ablation timescales as short as $\tau_\mathrm{abl} \equiv  m/\dot{m}_\mathrm{abl} \lesssim \mathrm{hr}$. The planetesimals captured in the retrograde directions both reached our imposed cut-off radius of $10\,$m (at this small size, the objects couple with the gas and end up rotating in the prograde direction which can be more clearly seen in the bottom right panel ). However, far from the pericenter of the orbit, ablation rapidly becomes negligible for planetesimals orbiting on both prograde and retrograde orbits ($\tau_\mathrm{abl} \gtrsim 10^7$--$10^8\,$hr $\gg \Omega^{-1}_\mathrm{K}$).}\label{Orb_CPD}
\end{figure*}

\paragraph{Structure of the protoplanetary disk} When far from the planet (i.e, at distances $>0.2\, R_\mathrm{H}$), planetesimals interact with the gas of the PPD.
In this case, the gas density and orbital velocity are taken from the results of a 2D hydrodynamic simulation including a Jupiter mass planet, similarly to that used in \citet{Ro18}, where an isothermal equation of state has been used for the gas and an $\alpha$-viscosity \citep{SS73} with $\alpha=2\times10^{-3}$ was assumed. 
The surface density of the PPD ($\Sigma_\mathrm{PPD}$) was normalized so that the unperturbed surface density (that is, in the absence of the planet) at $1\,$au is $\sim300 \, \mathrm{g\,cm}^{-2}$, corresponding to a moderately evolved disk \citep[e.g.,][]{BJLM15}.
The resulting surface density is shown in Fig.\ref{SD_PPD}. 
To obtain the volume density $\rho_g$ of gas around a planetesimal, we assume vertical hydrostatic equilibrium,
\begin{equation}\label{rho_PPD}
	\rho_g = \frac{\Sigma_\mathrm{PPD}}{\sqrt{2\pi}H_\mathrm{PPD}}\exp\left(-\frac{z^2}{2H^2_\mathrm{PPD}}\right),
\end{equation}
where $H_\mathrm{PPD}$ is the scale height of the disk and is derived from the fact that the hydrodynamic simulations were performed assuming a constant aspect ratio of the disk $H_\mathrm{PPD}/R=0.05$, $R$ being the radial distance to the star in cylindrical coordinates.

\paragraph{Structure of the circum-planetary disk} Planetesimals experiencing a close encounter with the planet (i.e., at distances $\leqslant0.2\,R_\mathrm{H}$) interact with the CPD. 
The surface density of the CPD follows the power-law presented in the previous section with an index $\gamma=1.5$ and a total mass $M_\mathrm{CPD}=1.5\times10^{-3}M_\mathrm{p}$, yielding $\Sigma_\mathrm{out}\approx 2\times10^{2}\,\mathrm{g\,cm}^{-2}$.
This differs from the work of \citet{SO17} who scaled their analysis of the capture of planetesimals in CPDs to the gas starved model, assuming $\Sigma_\mathrm{out}=1\,\mathrm{g\,cm}^{-2}$. In such a low density disk, the largest planetesimals that can be efficiently captured have a size of $R_\mathrm{pl}\sim 3\,$km. 
However, primordial asteroids of the Solar System seem to record a birth size distribution of objects such that most of the mass was carried by large planetesimals \citep{MBNL09}.
This is also consistent with the results of numerical simulations of the formation of planetesimals through the streaming instability, showing that the mass of the planetesimal population is dominated by objects with a radius of $\sim$100$\,$km \citep{Jo+15,SALY16}.
For our adopted CPD mass, which is consistent with recent numerical simulations \citep{Sz17}, hundred-kilometer-sized planetesimals can readily be captured. 
This alleviates the need to invoke substantial collisional grinding of the initial planetesimal population to allow for their delivery to the giant planets' CPDs \citep[e.g.,][]{Es+09,MEC10}.
We assume vertical hydrostatic equilibrium to obtain the volume density as a function of height in the disk.
Since the aspect ratio of the CPD is expected to vary only slightly with respect to distance from the planet in the inner portions of the disk, we here assume an average and constant aspect ratio $h_\mathrm{g}=0.06$, so that the results presented in this section are not particularly tied to a particular (viscous or passive) CPD model.
We nevertheless verified that similar results are obtained when considering a broken power-law such as presented in Section~\ref{CPD_model}.

\begin{figure*}
	\includegraphics[width=\linewidth]{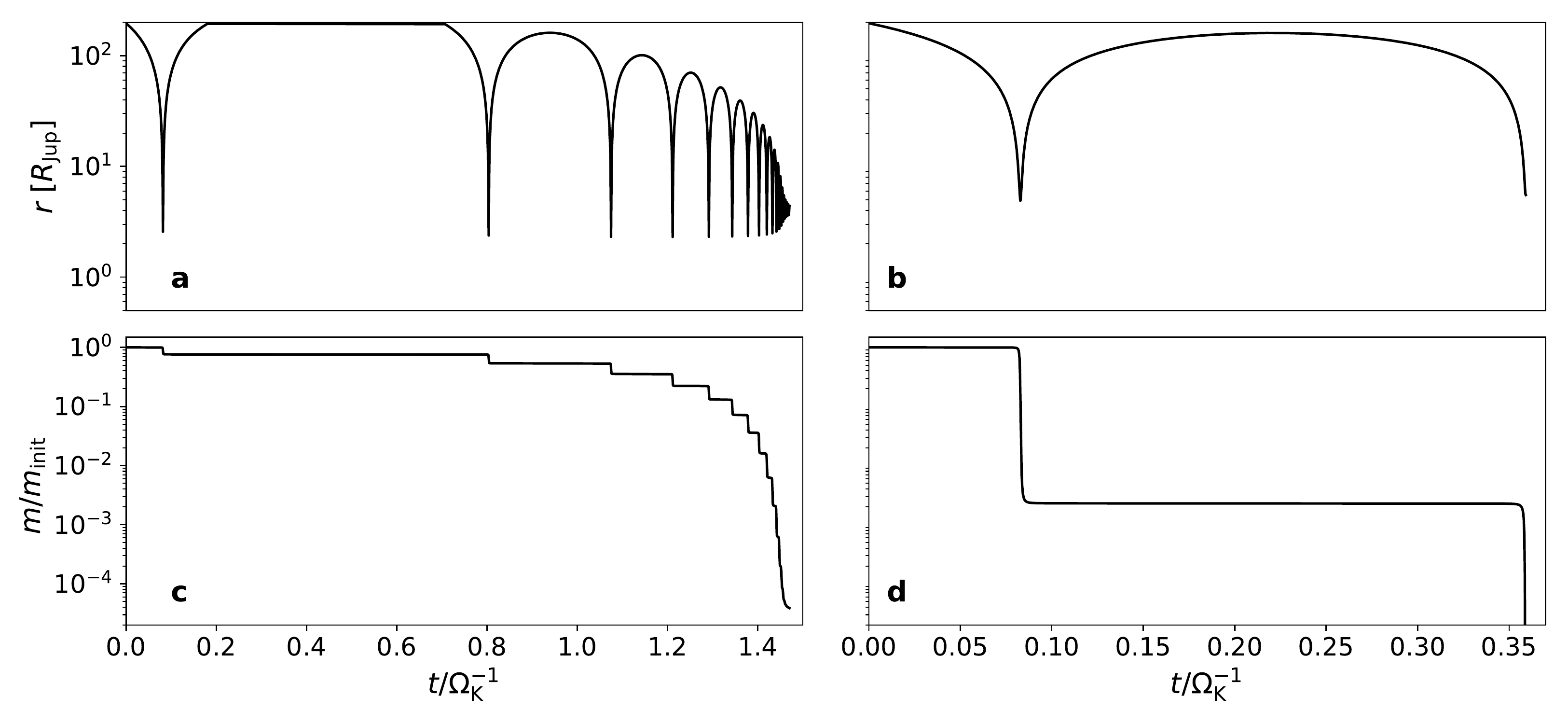}
	\caption{\textit{Top.} Evolution of the radial distance from the planet of a planetesimal captured in the prograde direction (panel a), and the retrograde direction (panel b), with an initial size $R_\mathrm{pl}=100\,$km, as a function of time (normalized by the inverse of the orbital angular frequency of the planet $\Omega_\mathrm{K}$). \textit{Bottom}. Corresponding evolution of the mass of the planetesimal due to ablation in the case of prograde (panel c) and retrograde (panel d) capture. Ablation essentially occurs at closest approach from the planet due to the strong frictional heating in the dense inner parts of the CPD.}\label{mass_evo}
\end{figure*}

\subsection{Results}

Here we present the results of the orbital integration of the planetesimals that were initially released on heliocentric orbits within the  feeding zone of the giant planet. 
Planetesimals found on bound orbits around the planet with a (planetocentric) semimajor axis $\leqslant 0.05\, R_\mathrm{H}$ ($\sim40\,R_\mathrm{Jup}$) and an eccentricity $<0.1$ were removed from the simulation to save computation time. For the range of planetesimal sizes and disk densities explored here, we find that our eccentricity criterion is generally the more restrictive, i.e., the semimajor axis of most captured planetesimals shrinks to values $\leqslant 0.05\, R_\mathrm{H}$ before their eccentricities are damped to values $<0.1$.
As shown below, we find that objects that have reached our removal criterion should not contribute further significant amounts of dust through ablation as they have either already lost most of their initial mass, or are found on more distant orbits and experience little frictional heating. 
We also imposed a minimum cut-off radius of $10\,$m after which point further ablation was not considered. Since we investigated the evolution of large planetesimals, objects that have reached the cut-off radius have a negligible mass ($m_\mathrm{cut-off}/m_\mathrm{init}=10^{-12}$--$10^{-6}$ when $R_\mathrm{pl}=100$--$1\,$km initially). 

\begin{figure}
	\includegraphics[width=\linewidth]{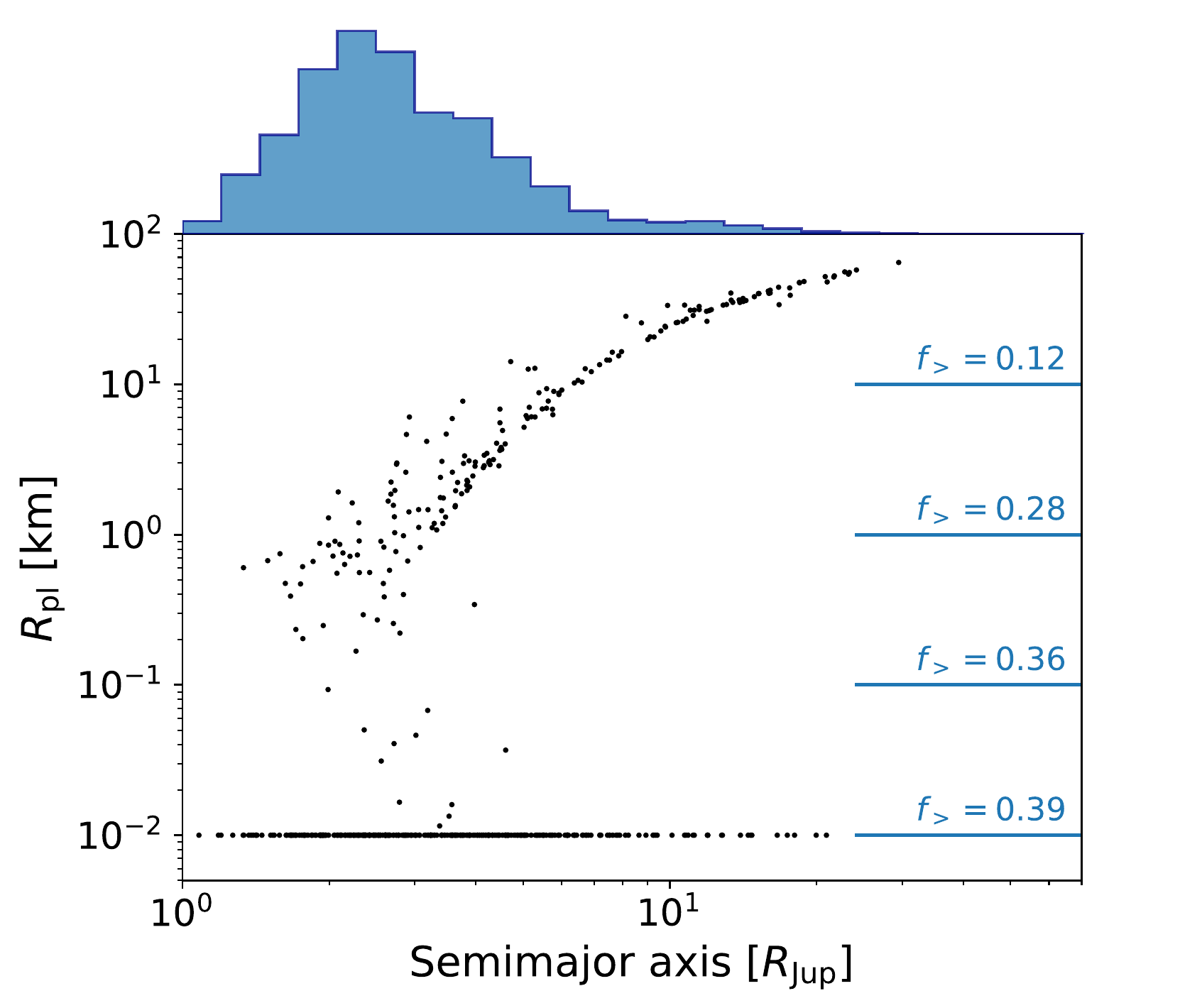}
	\caption{Final radius versus final semimajor axis of planetesimals in the circum-planetary disk (i.e., at the time of circularization of the orbit according to our criteria, $\tilde{a}\leqslant0.05 \, R_\mathrm{H} \sim 40\,R_\mathrm{Jup}$ and $\tilde{e}\leqslant 0.1$). Each planetesimal started with an initial radius $R_\mathrm{pl}=100\,$km at the beginning of the simulation. On the right, the fraction of objects with a size $R_\mathrm{pl}>R_0$, defined as $f_>=N(R_\mathrm{pl}>R_0)/N_\mathrm{tot}$, is reported for different threshold sizes $R_0$ as indicated by the horizontal blue lines. The histogram at the top shows the corresponding distribution of the final semimajor axis of the planetesimals. The vast majority of planetesimals are captured at distances $\lesssim 10 \,R_\mathrm{Jup}$ and are strongly, if not completely, ablated.} \label{a_Rpl}
\end{figure}

\subsubsection{Distribution of ablated material}\label{dist_abl}

Planetesimals are efficiently ablated only when they experience a very close encounter with the planet.
The ablation is mainly driven by the strong heating provided by friction with the gas, whereas the temperature of the CPD has a minor effect.
If we consider for example a planetesimal approaching the planet at the local escape velocity, in the case of a prograde motion with respect to the rotation of gas in the CPD, the relative velocity between the gas and the planetesimal would be $v_\mathrm{rel}=(\sqrt{2}-1)v_\mathrm{K}$, or $v_\mathrm{rel}=(\sqrt{2}+1)v_\mathrm{K}$ in the case of a retrograde approach. 
For a closest approach at a distance $r_\mathrm{CA}=10\,R_\mathrm{Jup}$, $\rho_\mathrm{g}\approx 10^{-6}\, \mathrm{g\,cm}^{-3}$.
Using these values to estimate the frictional heating (second term on the r.h.s of eq.~\ref{Tsurf}), we find it to be about five orders of magnitude larger than the heating provided by radiation at the local temperature of the CPD ($T_\mathrm{d} \approx 190 \,$K).
In fact, the temperature of the CPD should have been $T_\mathrm{d} \gtrsim 3000\,$K to provide a source of energy comparable to that of gas drag.

As a consequence, ablation mostly occurs in the densest inner parts of the CPD. 
This is illustrated in Fig.~\ref{Orb_CPD} which shows example trajectories of planetesimals with an initial radius $R_\mathrm{pl}=100\,$km that are captured by the planet in both prograde and retrograde directions.
Planetesimals captured on retrograde orbits have a larger relative velocity with respect to the gas than the planetesimals that are captured in the prograde directions. They are therefore more strongly heated and ablated.

Figure~\ref{Orb_CPD} reveals that planetesimals captured in the prograde direction have ablation timescales $\tau_\mathrm{abl} \equiv m/\dot{m}_\mathrm{abl} \lesssim 10\,$hr (where $m=m(t)$ is the mass of the planetesimal) near their closest approach (top left panel of Fig.~\ref{Orb_CPD}), whereas planetesimals captured in the retrograde direction have ablation timescales $\tau_\mathrm{abl} \lesssim \mathrm{hr}$ (bottom left panel of Fig.~\ref{Orb_CPD}).
These ablation timescales are to be compared with the time 'spent' at the closest approach,
\begin{equation}\label{t_CA}
	t_\mathrm{CA} \sim \frac{r_\mathrm{CA}}{v_\mathrm{peri}} \approx 3.5 \left(\frac{r_\mathrm{CA}}{5\,R_\mathrm{Jup}} \right)^{3/2}\, \mathrm{hr},
\end{equation}
where we have followed \citet{TO10} in assuming that the path length of the orbit can be approximated by the distance at closest approach $r_\mathrm{CA}$, and we have used $v_\mathrm{peri}\approx \sqrt{2}v_\mathrm{K}$ \citep[e.g.,][]{FOTS13} which is a good approximation when $\tilde{e}\approx 1$ (here $\tilde{e}$ is the planetocentric eccentricity of the orbit of a planetesimal and is indeed close to unity right after capture).

This simple estimate suggests that a large fraction of the mass of a planetesimal can be ablated during a single passage at pericenter for either prograde or retrograde orbits.
Figure~\ref{mass_evo} shows the time evolution of the distance to the planet and mass of a planetesimal captured in the prograde and retrograde direction, respectively, and confirm this suspicion. 
From close inspection at Fig.~\ref{mass_evo}, we find that the planetesimal captured in the prograde direction has a fraction of about 40\% of its initial mass ablated after its first close encounter with the planet (i.e., upon capture) whereas the planetesimal captured in the retrograde direction has been almost entirely ablated.
It is evident from these figures that ablation occurs primarily near the closest approach to the planet and is otherwise negligible (which could also be inferred from Fig.~\ref{Orb_CPD}).

It is interesting to note that by the time the two planetesimals shown on Fig.~\ref{mass_evo} are circularized (according to our criterion), they have been almost completely ablated. Their final masses are fractions of only $\sim$10$^{-5}$ and $10^{-9}$ that of their initial mass. 
We find that this is actually the fate of the vast majority of captured planetesimals. 
We report on Fig.~\ref{a_Rpl} the final radii and semimajor axes (i.e., at the time of circularization) of planetesimals that have been captured by the planet.
More than 60\% of the captured objects have been ablated down to our imposed cut-off radius of $10\,$m. Those are mainly planetesimals that were captured on initially retrograde orbits.
Objects that are captured in the prograde direction may 'survive' ablation, with planetesimals captured at wider orbital distances experiencing less ablation (a typical example is the object shown on the top right panel of Fig.~\ref{Orb_CPD}).
We find that about 10\% of the captured objects can remain as large objects in the CPD with sizes $R_\mathrm{pl}>10\,$km and they all have a (planetocentric) semimajor axis $\tilde{a} \gtrsim 10\,R_\mathrm{Jup}$.
Although these objects cannot be deemed as completely ablated, they have lost a large fraction of their mass through ablation.
Even the largest remaining planetesimal in our simulation ($R_\mathrm{pl}\approx 65\,$km) has lost more than half of its initial mass.

Due to the fact that ablation is efficient only at very close distances from the planet (see Figs.~\ref{Orb_CPD}, \ref{mass_evo}), the main contributors to the supply of material are planetesimals that are captured by the planet or directly collide with it. 
Planetesimals that merely fly-by the planet and the CPD deliver only a negligible amount of mass through ablation.

\begin{figure}
	\includegraphics[width=\linewidth]{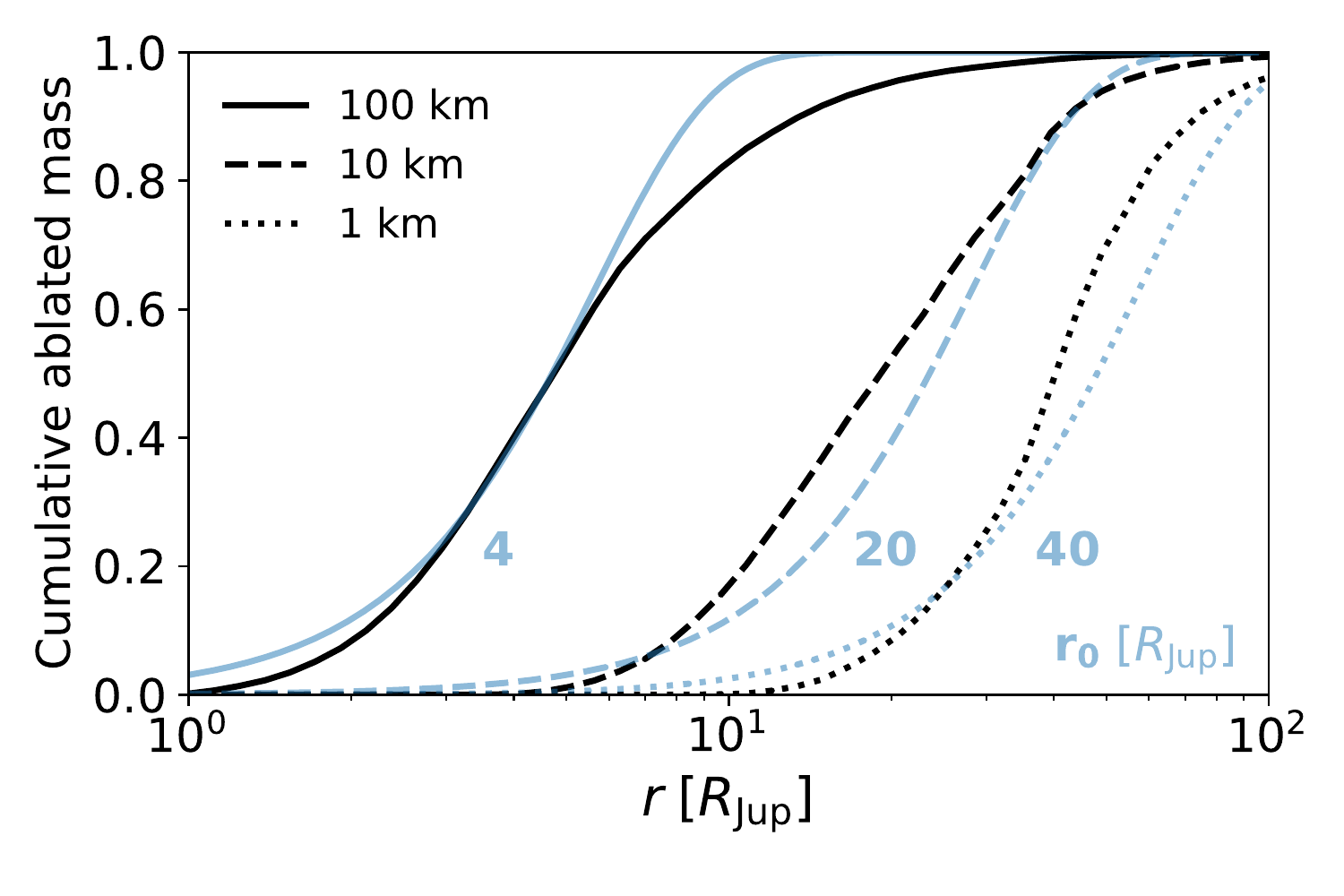}
	\caption{Cumulative mass distribution of material ablated off of planetesimals with initial sizes of $100\,$km (solid black line), $10\,$km (dashed black line) and $1\,$km (dotted black line) as a function of the radial distance from the planet. Smaller planetesimals are captured at wider distances and deliver material outward compared to the larger objects. The light blue curves correspond to cumulative Rayleigh distributions, $F(r;r_0)=1-\exp\left[-r^2/(2r^2_0)\right]$, that we use as approximate fits to the distribution of ablated material. The number to the right of each blue curve corresponds to the value of the parameter $r_0$ of the Rayleigh distribution.}\label{abl_distrib}
\end{figure}

We show in Fig.~\ref{abl_distrib} the distribution of ablated material as a function of the distance from the planet in terms of the cumulative ablated mass for planetesimals with three different initial sizes, $R_\mathrm{pl}= 1$, 10 and $100\,$km.
To construct these curves, we divided the CPD in logarithmic bins and added the mass ablated off of planetesimals in each bin at any time. 
We then considered the (normalized) cumulative of this mass distribution as it appears much less noisy.
We stress that these curves indicate where the mass is deposited on average by planetesimals but they do not represent the distribution of material in the CPD at any particular time (such considerations are presented in Sec.~\ref{Dust_evol}).
We find that the cumulative ablated mass in the CPD can be approximately fitted by a function of the form
\begin{equation}\label{Ray_fit}
	F(r;r_0) = 1 - e^{-r^2/(2r^2_0)},
\end{equation}
which actually corresponds to a cumulative distribution function of a Rayleigh distribution, where $r$ is the distance to the planet and $r_0$ is a scale parameter. 
Light blue curves on Fig.~\ref{abl_distrib} were obtained using eq.~\ref{Ray_fit} with different values of $r_0$ as indicated to the right of each curve.
Although the fits are far from perfect, they reproduce the 'S'-shaped distribution of material and we show in Sec.~\ref{Dust_evol} how eq.~\ref{Ray_fit} can be related to some physical quantities, allowing us to construct a simple model of the evolution of solids in the CPD.

\subsubsection{Accretion rate onto the circum-planetary disk}

Keeping track of the total mass being ablated within the circum-planetary disk, we find that after 100 orbital periods of the planet, it amounts to $\sim$23\% of the total mass of planetesimals initially in the feeding zone of the planet in the case of $100\,$km sized objects.
If the initial surface density of planetesimals in the feeding zone would correspond to $\Sigma_\mathrm{pl}=1\,\mathrm{g\,cm}^{-2}$, the average accretion rate of solids onto the CPD due to ablation of planetesimals would be $\approx4.5\times10^{-5}\,M_\oplus \,\mathrm{yr}^{-1}$.
At the same time, the feeding zone has been emptied by $\approx$65\%, which means that it would be cleaned up in $\sim$10$^4$ years, irrespective of the initial surface density of planetesimals. 

This is problematic if the satellites of the giant planets must form on timescales of $10^5$--$10^6$ years \citep[e.g.,][]{BC08}.
However, as shown in \citet{Ro18}, the feeding zones of giant planets could be fed themselves as neighbouring planets form and destabilize nearby planetesimals, thus allowing the delivery of material to the CPDs on longer timescales. 

\section{Evolution of pebbles in the circum-planetary disk}\label{Dust_evol}

\begin{figure}
	\includegraphics[width=\linewidth]{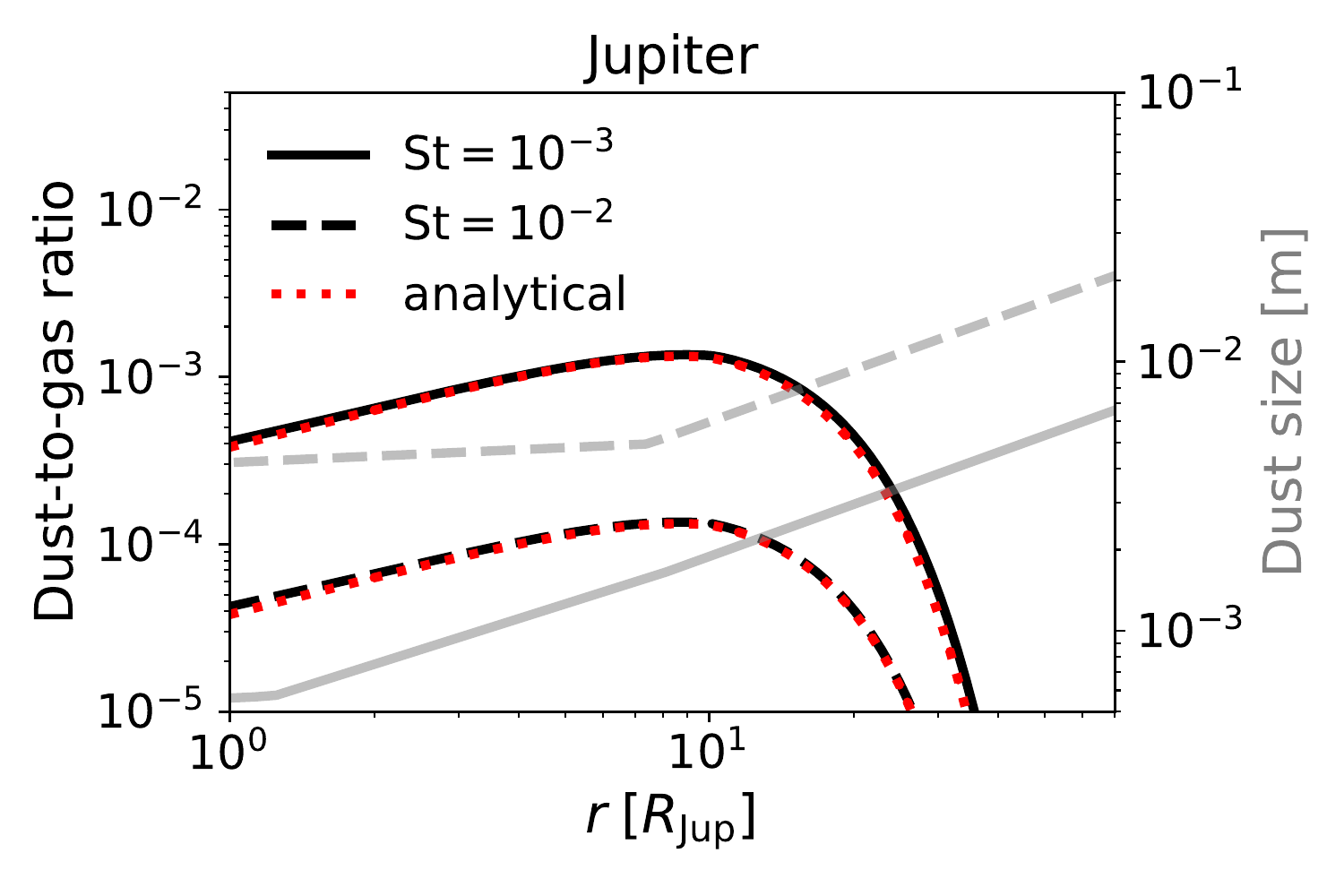}
	\includegraphics[width=\linewidth]{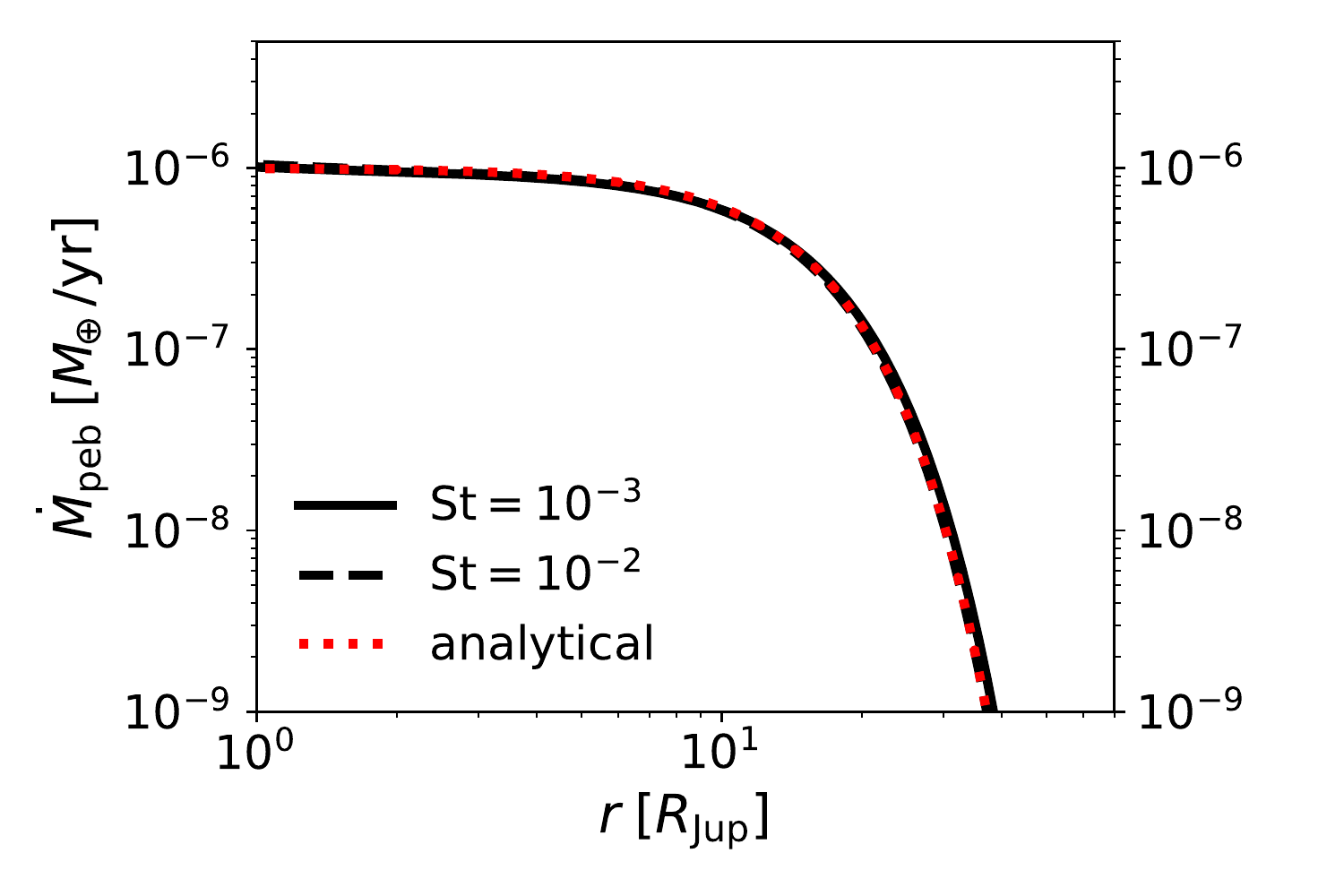}
	\caption{ \textit{Top:} Dust-to-gas mass ratio in the circum-planetary disk of a Jupiter-mass planet for pebbles Stokes number of $10^{-3}$ (black solid lines) and $10^{-2}$ (black dashed lines), assuming a characteristic radius of deposition of ablated material $r_0=10\,R_\mathrm{Jup}$ and a global supply rate $\dot{M}_0=1\,M_\oplus\,\mathrm{Myr}^{-1}$, obtained through numerical integration of the advection-diffusion equation of the evolution of the dust surface density after $5\times10^3\,$years of integration. 
	The red dotted curves show the corresponding analytical steady-state solutions given by eq.~(\ref{dust_to_gas}).
	The gray curves show the physical size of pebbles as a function of distance from the planet corresponding to Stokes number of $10^{-3}$ (solid) and $10^{-2}$ (dashed). \textit{Bottom:} Radial flux of pebbles through the CPD for the two considered Stokes number of the pebbles (overlapping black curves), along with the steady-state solution where $\dot{M}_\mathrm{peb}=\dot{M}_0 e^{-r^2/(2r^2_0)}$ (red dotted curve).}\label{dustCPD}
\end{figure}

The ablation of planetesimals in the CPD provides an important source of small dust grains from which satellites can accrete (this process is discussed in Section~\ref{accretion}).
We thus construct here a simple parametrization of the dusty component of a CPD fed by planetesimals ablation.

\paragraph{Dust supply} In the previous section, we have found that the cumulative ablated mass within the CPD could be roughly approximated by a cumulative Rayleigh distribution (eq.~\ref{Ray_fit}), corresponding to the following integral
\begin{equation}\label{Rayleigh}
	F(r;r_0) = 1 - e^{-r^2/(2r^2_0)} = \int^r_0 \frac{x}{r^2_0}e^{-x^2/(2r^2_0)} dx.
\end{equation}
Using physical quantities, the cumulative ablated mass in the CPD can be expressed as
\begin{equation}\label{M_abl}
	\frac{M(r)}{M_\mathrm{tot}} = \frac{\int^r_0 2\pi r \Sigma(r) dr}{M_\mathrm{tot}},
\end{equation}
where $M(r)$ is the total mass that has been deposited through ablation up to the distance $r$ from the planet, $\Sigma(r)$ is then the surface density distribution of the ablated mass and $M_\mathrm{tot}$ is the total mass that has been deposited in the CPD through ablation. By identification of the terms in eqs.~\ref{Rayleigh} and \ref{M_abl}, we find
\begin{equation}\label{Sigma_abl}
	\Sigma(r) = \frac{M_\mathrm{tot}}{2\pi r^2_0}  e^{-r^2/(2r^2_0)}.
\end{equation}
The above expression yields the distribution of material deposited by ablation as a function of the distance from the planet.
Assuming that mass is globally deposited onto the CPD at a rate $\dot{M}_0$, the source of dust provided by ablation can then be expressed as
\begin{equation}\label{dotSigma}
	\dot{\Sigma}_\mathrm{d} = \frac{\dot{M}_0}{2\pi r^2_0} e^{-r^2/(2r^2_0)}.
\end{equation}
The value of $r_0$, which controls where the ablated mass is deposited, should depend on the structure of the CPD and the planetesimal size distribution in the feeding zone of the giant planet (see Sect.~\ref{dist_abl} and Fig.~\ref{abl_distrib}) which are rather uncertain. 
We  thus treat it as a free parameter, with plausible values of the order of 10 planetary radii considering our results and the current radial extent of the satellite systems of Jupiter and Saturn.

\paragraph{Quasi-steady-state analytical solution} The dust released through planetesimal ablation is not bound to remain where it has been deposited.
The dust grains should rapidly grow and start to radially drift inward.
The dust growth timescale is proportional to the orbital timescale and depends on the dust-to-gas mass ratio in the disk \citep[e.g.,][]{BKE12}:
\begin{equation}
 \tau_\mathrm{growth} \equiv \frac{s_\bullet}{\dot{s}_\bullet} \approx \frac{1}{\varepsilon_\mathrm{d}\Omega_\mathrm{K}} \approx 1.65\times10^{-3} \varepsilon^{-1}_\mathrm{d} \left(\frac{r}{10\,R_\mathrm{jup}} \right)^{3/2} \mathrm{yr},
\end{equation}
where $s_\bullet$ is the radius of the dust grains, $\varepsilon_\mathrm{d}=\Sigma_\mathrm{d}/\Sigma_\mathrm{g}$ is the dust-to-gas mass ratio, and the numerical estimate is for a Jupiter-mass planet.
The radial drift velocity of dust, on the other hand, is a function of the aerodynamic coupling of the dust particles with the gas, which is expressed through their Stokes number, $\mathrm{St}=t_\mathrm{s}\Omega_\mathrm{K}$, and is given by
\begin{equation}
	v_r = -\frac{2\mathrm{St}}{1+\mathrm{St}^2}\eta v_\mathrm{K},
\end{equation}
where $v_\mathrm{K}$ is the keplerian orbital velocity around the planet and $\eta$ is a measure of the gas pressure support of the CPD,
\begin{equation}
	\eta = -\frac{1}{2} \frac{\mathrm{d} \ln P}{\mathrm{d}\ln r} h^2_\mathrm{g},
\end{equation}
 with $P$ the pressure in the midplane of the CPD. 
This yields the radial drift timescale of dust,
\begin{equation}
	\begin{split}
	\tau_\mathrm{drift} \equiv \left|\frac{r}{v_r} \right| & \approx \frac{1}{2\eta\mathrm{St}}\Omega^{-1}_\mathrm{K} \\
	{} & \approx 13 \left(\frac{10^{-2}}{\mathrm{St}} \right) \left( \frac{0.06}{h_\mathrm{g}}\right)^2 \left(\frac{r}{10\,R_\mathrm{Jup}} \right)^{3/2} \mathrm{yr}.\\
	\end{split}
\end{equation}

Given these short growth and radial drift timescales, it is likely that the dust distribution in the CPD rapidly reaches a quasi-steady-state regulated by the rate at which dust is supplied by ablation. This should hold as long as $\dot{M}_0$ does not significantly vary over timescales $\gg \tau_\mathrm{drift}$.
The steady-state solution can be obtained by considering that the flux of pebbles through the disk at any given distance $r$ must be equal to the total flux deposited exterior to $r$ by ablation, translating to 
\begin{equation}\label{flux}
	2\pi r v_r \Sigma_\mathrm{d} = \int_r^{+\infty}\! 2\pi r \dot{\Sigma}_\mathrm{d} dr = \dot{M}_0 e^{-r^2/(2r^2_0)}.
\end{equation}
When $\mathrm{St}\ll 1$, the drift velocity is $v_r \approx 2 \mathrm{St} \eta v_\mathrm{K}$, which, using eq.~\ref{flux}, yields the following expression of the dust-to-gas mass ratio in the CPD,
 \begin{equation}\label{dust_to_gas}
	\varepsilon_\mathrm{d} \approx \frac{\dot{M}_0 e^{-r^2/(2r^2_0)}}{4\pi r \mathrm{St} \eta v_\mathrm{K} \Sigma_\mathrm{g}}.
\end{equation} 

In Fig.~\ref{dustCPD}, we show a comparison of the dust-to-gas mass ratio and pebble flux through the CPD of a Jupiter-mass planet, obtained by numerically solving the radial advection-diffusion equation of the dust surface density \citep[e.g.,][]{BKE12}, with the analytical solutions given by eqs.~\ref{flux} and \ref{dust_to_gas} for different values of the Stokes number of the dust particles.
We have used here a characteristic deposition radius $r_0=10\,R_\mathrm{Jup}$ and a global supply rate $\dot{M}_0=1 \,M_\oplus\,\mathrm{Myr}^{-1}$ (equivalently $3\times10^{-9}\, M_\mathrm{p}\,\mathrm{yr}^{-1}$, for a Jovian-mass planet).
The numerical integration results were taken after $5\times10^3\,$years of evolution and effectively show a steady-state which agrees well with the analytical estimates. 
It is interesting to note that the flux of pebbles through the disk is indeed being regulated by the ablation supply rate and thus independent of the Stokes number of the particles (bottom panel of Fig.~\ref{dustCPD}).
The quasi-steady-state solution for the pebble flux and surface density equally applies to the case of the CPD of a Saturn-mass planet as the dust growth and drift timescales are equivalently short.

The top panel of Fig.~\ref{dustCPD} also shows the physical size of the dust grains in the CPD corresponding to the assumed Stokes number.
It is important to note that, in the CPD, the mean free path of the gas molecules is much shorter than the typical values encountered in protoplanetary disks. 
Thus, pebbles are found in the Stokes drag regime or even the non-linear drag regime closer to the planet (as indicated by the change of slope of the gray curves in the top panel of Fig.~\ref{dustCPD}), as opposed to the Epstein regime which is the most relevant under PPD conditions \citep[see][for a review of the different drag regimes]{Jo+14}.
The relation between the Stokes number of the particles and their physical sizes in the relevant regimes are presented in Appendix~\ref{App_dust}.

\begin{figure}
	\includegraphics[width=\linewidth]{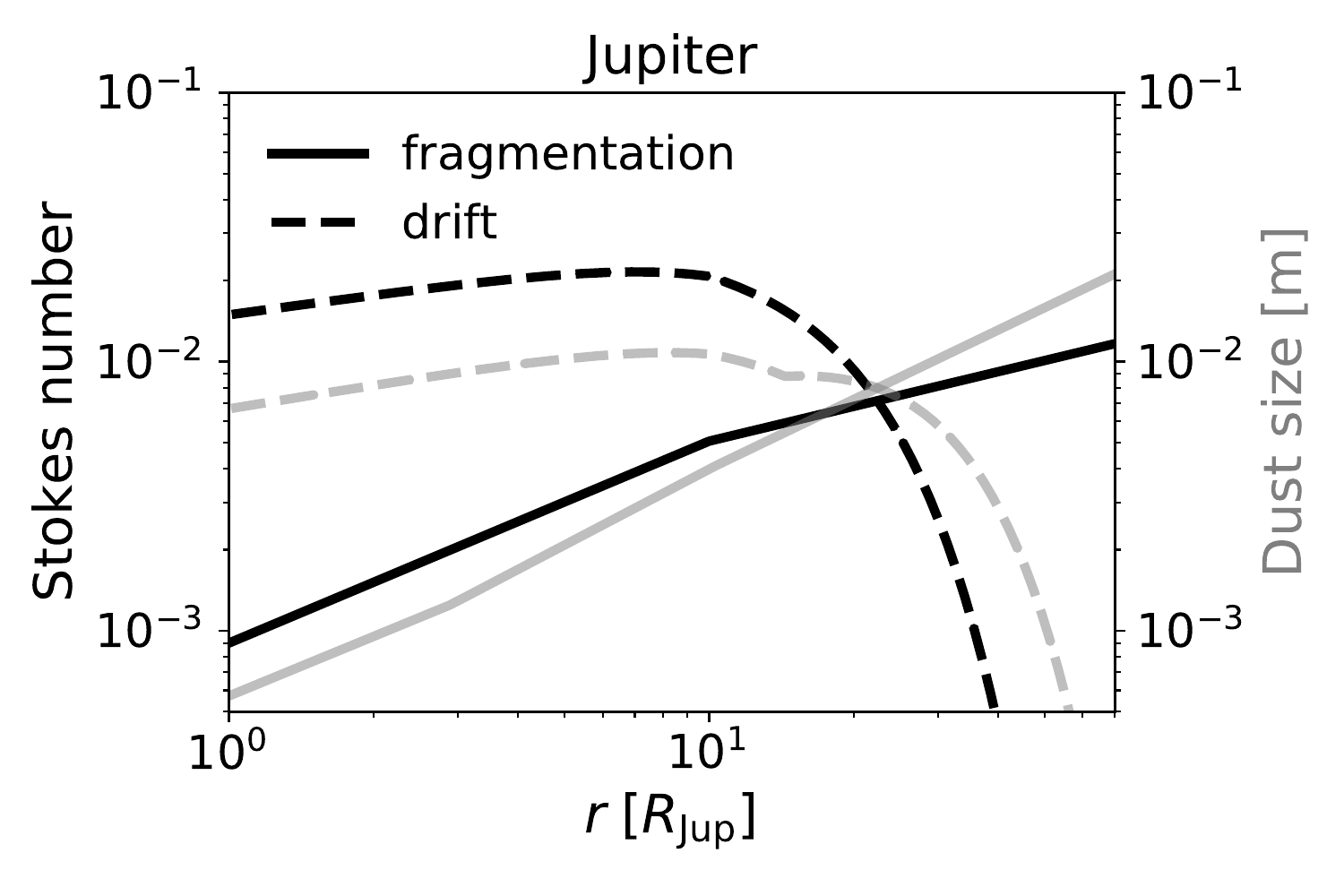}
	\includegraphics[width=\linewidth]{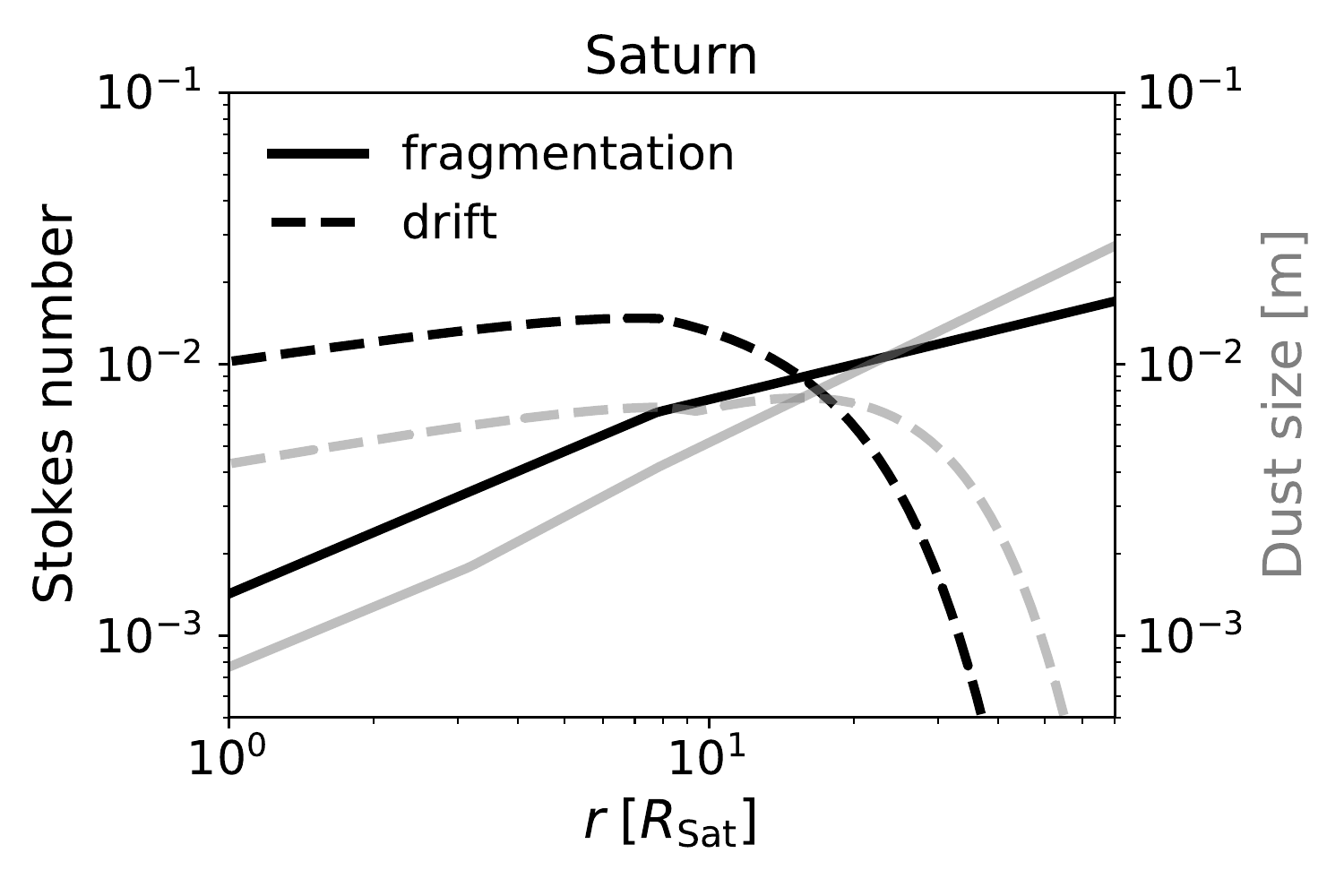}
	\caption{Comparison of the Stokes number of the dust particles in the CPDs of Jupiter (top) and Saturn (bottom) in the fragmentation limited regime (black solid lines), assuming $v_\mathrm{f}=1\,\mathrm{m\,s}^{-1}$ and $\alpha=10^{-4}$, and the drift limited regime (black dashed lines), assuming $\dot{M}_0 = 10^{-9}\, M_\mathrm{p}\,\mathrm{yr}^{-1}$ and $r_0 = 10\,R_\mathrm{Jup/Sat}$. The solid and dashed gray curves show the physical sizes of the dust grains in the fragmentation and drift limited regimes, respectively.}\label{Frag_Drift}
\end{figure}

\begin{figure*}
	\includegraphics[width=\linewidth]{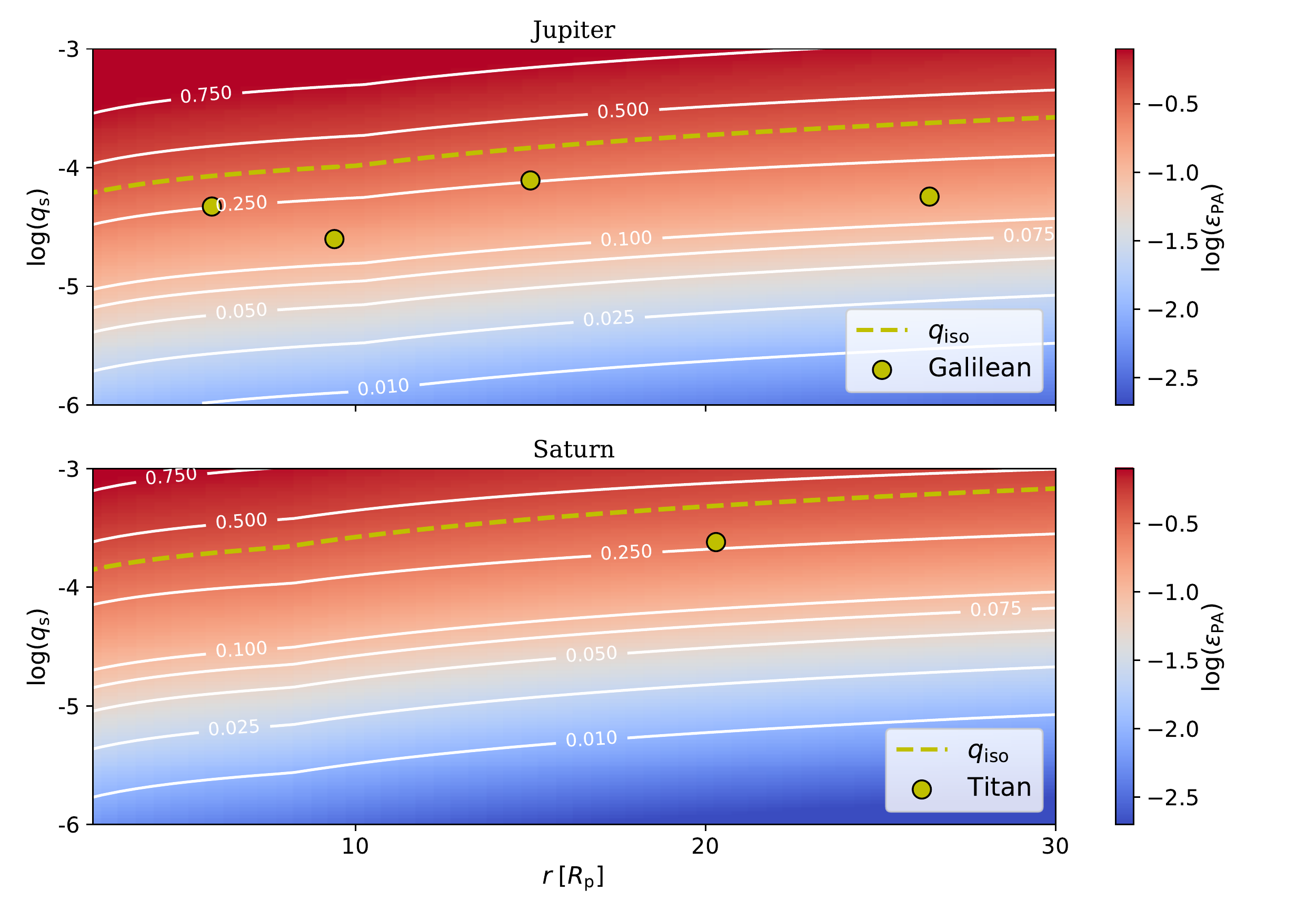}
	\caption{Pebble accretion efficiency as a function of distance from the planet and mass of the protosatellite ($q_\mathrm{s}=M_\mathrm{s}/M_\mathrm{p}$ is the ratio of the mass of the seed to that of the planet) in a CPD surrounding Jupiter (top panel) or Saturn (bottom panel). Here, pebbles with a constant Stokes number ($\mathrm{St} = 5\times10^{-3}$) were assumed. In both panels, yellow dots correspond to the current mass and orbital distances of the major satellites of Jupiter (Galilean moons), or Saturn (Titan). The yellow dashed line shows the pebble isolation mass as a function of the distance from the planet, which is the maximum mass a satellite can reach through pebble accretion.}\label{eff_PA_irr_visc}
\end{figure*}

\paragraph{Pebbles Stokes number} Although the flux of pebbles through the CPD does not depend on their Stokes number in the steady-state, the pebble accretion rate onto a growing moon does \citep[e.g.,][and Sect.~\ref{accretion}]{LJ12}. 
Dust particles can initially grow by coagulation, but this growth is limited by either the rapid radial drift of the pebbles or fragmenting collisions at high relative velocities \citep[e.g.,][]{BDH08}.
In the drift limited regime, pebbles can grow up to the point when their drift timescale becomes comparable to their growth timescale, which occurs at a Stokes number of \citep{BKE12}
\begin{equation}
 \mathrm{St_{drift}}=2\varepsilon_\mathrm{d} \eta^{-1}.
\end{equation}
Using the expression of the dust-to-gas mass ratio in the CPD given by eq.~\ref{dust_to_gas} yields
\begin{equation}
\mathrm{St_{drift}} \approx \left(\frac{\dot{M}_0e^{-r^2/(2r^2_0)}}{2\pi r \eta^2 v_\mathrm{K} \Sigma_\mathrm{g}} \right)^{1/2}.
\end{equation}
On the other hand, fragmentation occurs when the relative velocity between colliding particles becomes higher than a threshold fragmentation speed $v_\mathrm{f}$.
The corresponding Stokes number of the particles is then \citep{BKE12}
\begin{equation}
\mathrm{St_{frag}}=\frac{1}{3} \alpha^{-1}\left(\frac{v_\mathrm{f}}{c_\mathrm{g}} \right)^2,
\end{equation}
where $\alpha$ is the turbulent viscosity parameter.

The maximum Stokes number achieved in the fragmentation and drift limited regimes in the CPDs of Jupiter and Saturn\footnote{We assume that Saturn formed at $\approx$7$\,$au, which would roughly correspond to the location of the outer edge of the gap opened by Jupiter and a 3:2 mean motion resonance configuration between Jupiter and Saturn \citep[see, e.g.,][for a more detailed description of this scenario]{Ro18}. In this case, for Saturn's CPD, $r_\mathrm{out}\approx 164\,R_\mathrm{Sat}$ and $\Sigma_\mathrm{out}\approx 7.5 \times 10^2 \, \mathrm{kg\,m}^{-2}$.} are presented in Fig.~\ref{Frag_Drift} along with the corresponding particle sizes.
For both Jupiter and Saturn, we assume a dust mass supply of $\dot{M}_0=3\times10^{-9}\,M_\mathrm{p}\,\mathrm{yr}^{-1}$ and a characteristic deposition radius of 10 planetary radii when estimating the drift limited Stokes number of the particles.
As for fragmentation, we consider a turbulent viscosity parameter of $\alpha=10^{-4}$, in agreement with our assumption of low-viscosity, passively irradiated CPDs, and a threshold fragmentation speed $v_\mathrm{f}=1\,\mathrm{m\,s}^{-1}$ \citep[e.g.,][]{BW08}.
The resulting Stokes number and particle sizes are very similar for both planets.
The drift limit typically yields Stokes numbers of the order of $10^{-2}$, corresponding to roughly $\sim$cm sized pebbles, except at distances larger than a few $r_0$, where there is little dust deposition, which promotes particle drift over their growth.
Fragmentation limits the Stokes number of the pebbles to values that are typically lower than in the case of the drift regime (except, again, when far from $r_0$), with St in the range $10^{-3}$--$10^{-2}$ (corresponding to physical sizes of $\sim$1--10$\,$mm).
The growth of pebbles should thus mainly be limited by fragmentation in the regions where the massive regular satellites of Jupiter and Saturn are found and the moons would therefore have accreted from pebbles with small Stokes numbers ($\mathrm{St}\lesssim 10^{-2}$).

\section{Accretion of moons}\label{accretion}

Now that we have parametrized an ablation supplied circum-planetary disk, we turn to the discussion of the accretion of moons in such CPDs. 
We derive some characteristics of the accretion process in our proposed framework, with the aim of paving the way for future investigations using dedicated numerical simulations.

\begin{figure*}
	\centering
	\includegraphics[width=.8\linewidth]{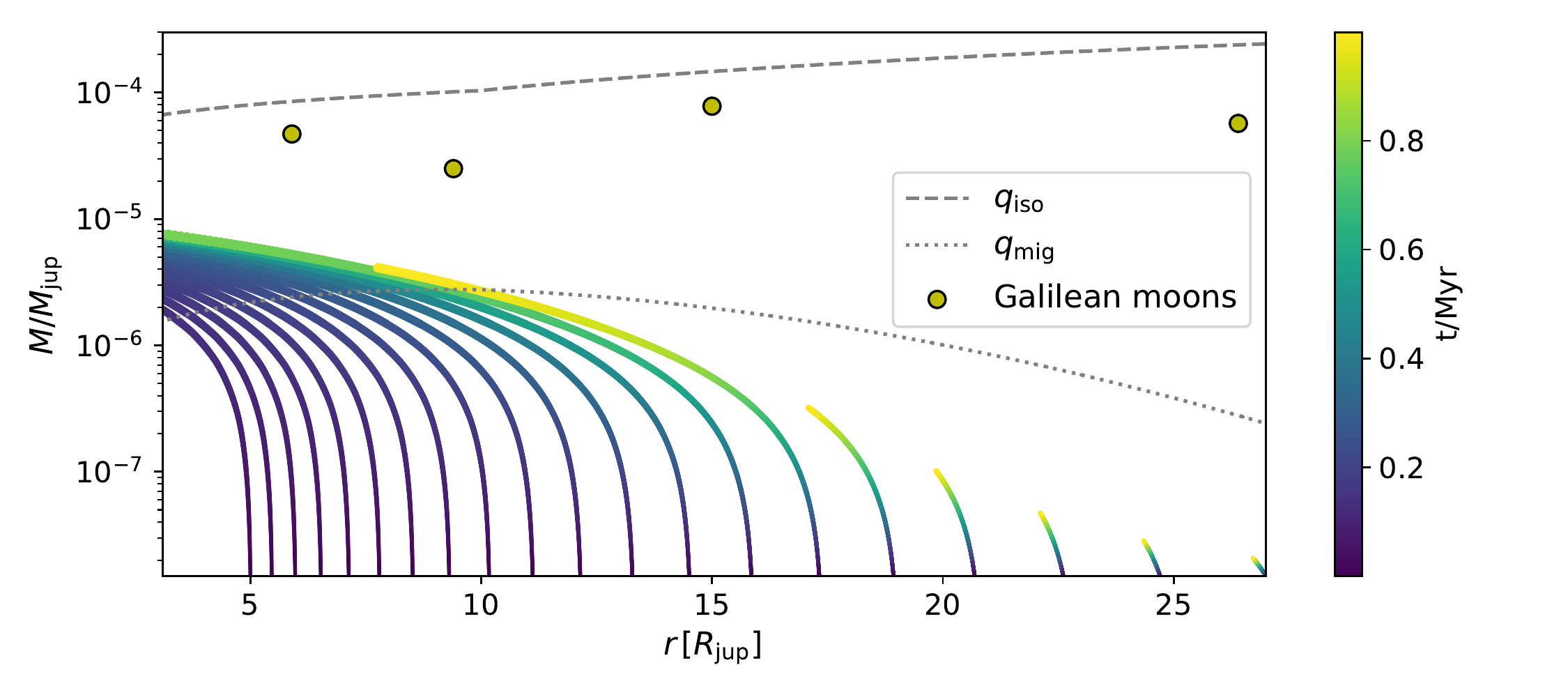}
	\caption{Growth tracks of pebble accreting and migrating protosatellites in the circum-jovian CPD assuming $\dot{M}_0=10^{-6}\,M_\oplus \, \mathrm{yr}^{-1}$, $r_0=10\,R_\mathrm{Jup}$, and pebbles with a constant Stokes number $\mathrm{St}=5\times10^{-3}$. Here, clearly, protosatellites cannot grow to masses comparable to that of the Galilean moons (marked as yellow dots) before reaching the inner edge of the disk.}\label{CPD_migration}
\end{figure*}

\subsection{Pebble accretion }

Once a protosatellite grows massive enough, it will be able to catch the radially drifting dust grains in a process known as pebble accretion \citep[see, e.g.,][]{LJ12,LJ14,Or17,JL17}.
Here we use the formalism of \citet{LO18} and \citet{OL18}, where the pebble accretion rate $\dot{M}_\mathrm{PA}$ is given by
\begin{equation}\label{Mdot_PA}
\dot{M}_\mathrm{PA} = \varepsilon_\mathrm{PA} \dot{M}_\mathrm{peb}.
\end{equation}
In the above expression, $\dot{M}_\mathrm{peb}$ represents the mass flux of pebbles in the CPD.
As we have shown in Section~\ref{Dust_evol}, the flux of pebbles should be at equilibrium with the supply of material through ablation of planetesimals, implying 
\begin{equation}\label{Mdot_peb}
\dot{M}_\mathrm{peb}=\dot{M}_0 e^{-r^2/2r^2_0}.
\end{equation}
The term $\varepsilon_\mathrm{PA}$ in Eq.~(\ref{Mdot_PA}) is the pebble accretion efficiency (i.e., the fraction of pebbles drifting past an accreting seed that is captured), for which the aforementioned authors provide formulae fitted to numerical simulations that we report in Appendix~\ref{App3}. 

Maps of the pebble accretion efficiency as a function of protosatellite mass and distance from the planet are shown in Fig.~\ref{eff_PA_irr_visc} in the case of the Jovian and Saturnian CPDs, assuming that pebbles have a constant Stokes number $\mathrm{St}=5\times10^{-3}$.
At mass ratios comparable to that of the Galilean moons and Titan ($q_\mathrm{s}\sim10^{-4}$), the pebble accretion efficiency is of the order of $\sim$25\%, and it is less than $\sim$1\% when $q_\mathrm{s} \lesssim 10^{-6}$.
This implies that only a small fraction of the available pebble flux will be accreted by the moons over the course of their formation and that more than the current mass of the satellite systems worth of pebbles must have been processed through the CPDs of the giant planets.
We discuss this and some implications in Sect.~\ref{peb_eff}.

\paragraph{Onset of pebble accretion} 
A necessary condition for pebble accretion to operate is that the timescale during which a pebble interacts with an accreting seed is comparable to (or shorter than) its stopping time \citep[see, e.g.,][]{Or17}. According to \citet{LO18}, this condition is satisfied when
\begin{equation}\label{q_PA}
	q_\mathrm{s} \gg \eta^3 \mathrm{St} \approx 1.25\times10^{-9} \left(\frac{\mathrm{St}}{5\times 10^{-3}}\right) \left( \frac{h_\mathrm{g}}{0.06}\right)^6.
\end{equation}
Around Jupiter, this corresponds to an object with a size of $\approx$83$\,$km.
Around Saturn, assuming a thicker aspect ratio $h_\mathrm{g}\sim 0.08$, this would correspond to a size of $\approx$98$\,$km.
The largest captured objects within the CPD could thus readily accrete pebbles, although they might initially grow by accreting other large captured planetesimals as these latter would dominate at distances beyond the characteristic ablation deposition radius $r_0$ (see Fig.~\ref{a_Rpl}).

\begin{figure*}
	\includegraphics[width=\linewidth]{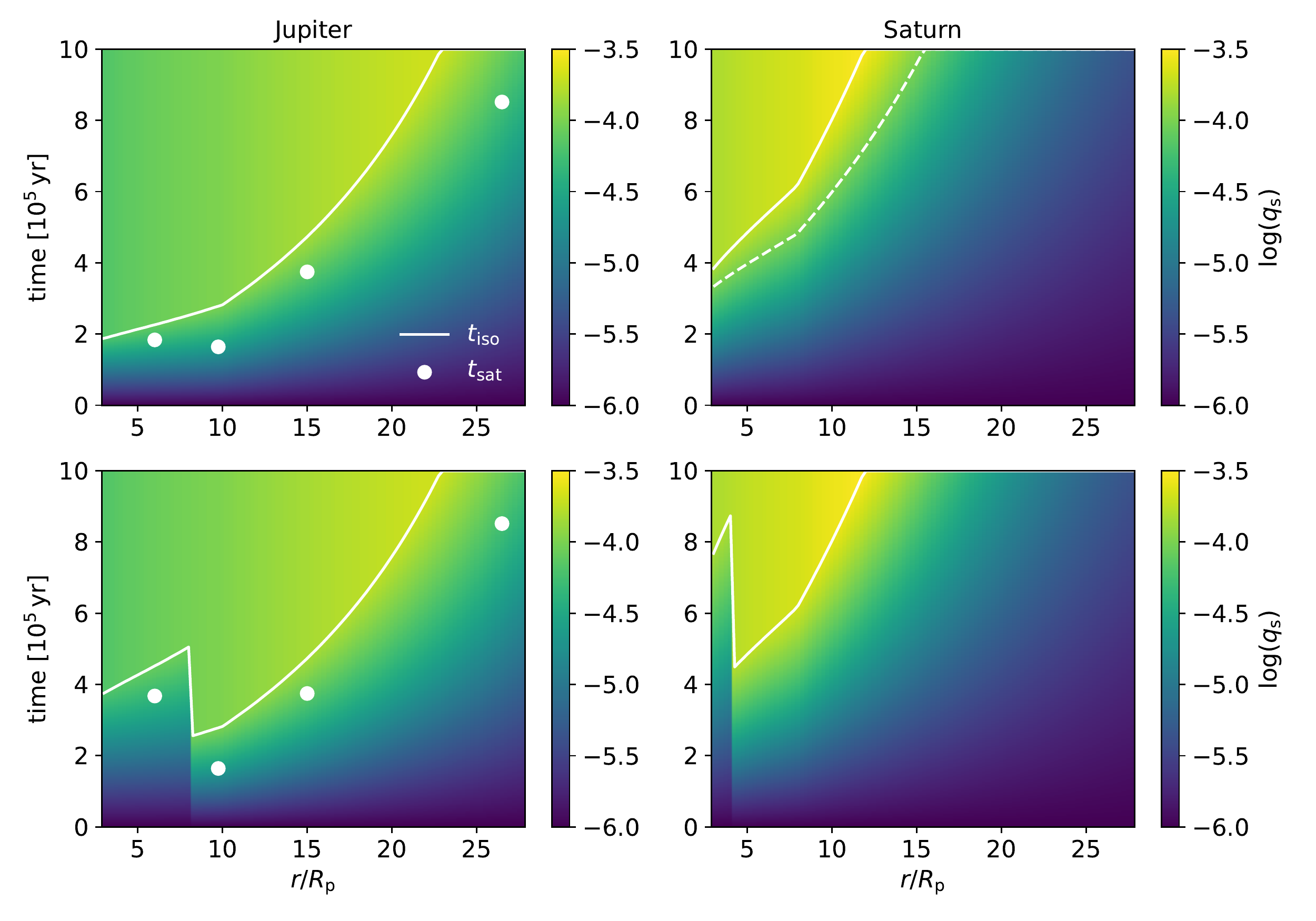}
	\caption{Growth maps of protosatellites through pebble accretion as a function of distance from their parent planet and time, in the case of Jovian (left panels) and Saturnian (right panels) circum-planetary disks. A fixed Stokes number, $\mathrm{St}=5\times10^{-3}$, has been assumed, as well as a characteristic deposition radius of ablated material $r_0=20\,R_\mathrm{Jup/Sat}$ and a mass accretion rate $\dot{M}_0=3\times10^{-9}\,M_\mathrm{p}\,\mathrm{yr}^{-1}$. Top panels show cases where the flux of pebbles through the disk is $\dot{M}_\mathrm{peb}=\dot{M}_0 \exp[-r^2/(2r^2_0)]$. The bottom panels include the effect of a snowline, assuming that the flux of pebbles is halved inside of the snowline due to the sublimation of water ice. The white line in each panel indicates the time needed to reach the pebble isolation mass at a given radial distance, whereas the white dots in the top and bottom left panels mark the time at which the mass of the Galilean moons is reached at their present location (but see Fig.~\ref{Growth_tracks} for actual growth tracks of Galilean-like moons). The time needed to reach the mass of Titan at its present location ($\sim$20$\,R_\mathrm{Sat}$) exceeds 1 My for the assumed parameters, and we argue in Sect.~\ref{Dyn_Insta} that Titan may have formed from the merging of lower mass moons following a dynamical instability.}\label{Growth_map}
\end{figure*}

\paragraph{Pebble isolation mass} An important aspect of pebble accretion is that it has an end. 
As an object grows more and more massive, it will start to significantly perturb the distribution of gas around it and carve an initially shallow gap.
The outer edge of the gap acts as a barrier to the drift of pebbles which will then be unable to reach the accreting seed, hence terminating pebble accretion \citep{MN12,LJM14,Bi+18}.
 Using 3D hydrodynamical simulations, \citet{LJM14} derived the mass at which pebble accretion ends, the so-called isolation mass. 
Using their scaling relation, we find that, in the CPD, pebble isolation occurs at
\begin{equation}\label{q_iso}
	q_\mathrm{iso} = 6\times10^{-5} \left(\frac{h_\mathrm{g}}{0.05} \right)^3.
\end{equation}

If the final mass of satellites is set by the pebble isolation mass, eq.~(\ref{q_iso}) implies that the aspect ratio of the jovian CPD was $h_\mathrm{g}\approx 0.055$ at the time and place Ganymede finished its growth, whereas it should have been $h_\mathrm{g}\approx 0.08$ around Saturn where and when Titan has formed.
The pebble isolation mass as a function of distance from the planet in the CPDs of Jupiter and Saturn is drawn as a yellow dashed line in Fig.~\ref{eff_PA_irr_visc}.

\subsection{Type-I migration}

As a protosatellite grows more and more massive, it will start to migrate inward due to tidal interactions with the gaseous disk (so-called type-I migration). 
The type-I migration rate is classically expressed as
\begin{equation}\label{typI}
	\dot{r} = - k_\mathrm{mig} q_\mathrm{s} \frac{\Sigma_\mathrm{g} r^2}{M_\mathrm{p}} h^{-2}_\mathrm{g} v_\mathrm{K},
\end{equation}
where $k_\mathrm{mig}$ is a constant that depends on the temperature and surface density gradients of the disk. 
Since the type-I migration rate of a protosatellite is directly proportional to its mass, the induced orbital decay is a threat to the growth of moons \citep[see][for an investigation of this issue in the case of planets]{JIB19}.

A rough estimate of the mass a protosatellite can reach before migration will dominate its evolution can be inferred by equating its growth timescale, $\tau_\mathrm{gr}\equiv M_\mathrm{s}/\dot{M}_\mathrm{PA}$, with its type-I migration timescale, $\tau_\mathrm{I}\equiv r/\dot{r}$ \citep[e.g.,][]{CW02}.
In the 3D pebble accretion regime, this migration limited mass is
\begin{equation}\label{q_mig}
\begin{split}
	q_\mathrm{mig}  =& \frac{0.39}{k_\mathrm{mig}} \frac{h^2_\mathrm{g}}{\eta h_\mathrm{peb}} \frac{\dot{M}_\mathrm{peb} \Omega_\mathrm{K}}{\Sigma_\mathrm{g} r^{2}} \approx 3.85\times10^{-6} e^{-r^2/(2r^2_0)} k^{-1}_\mathrm{mig} \\
	{} & \times \left(\frac{r}{r_0} \right)^{5/7} \left(\frac{\dot{M}_0}{3\times10^{-9} M_\mathrm{p}\,\mathrm{yr}^{-1}} \right) \left(\frac{M_\mathrm{CPD}}{1.5\times10^{-3}M_\mathrm{p}} \right)^{-1} \\
	{} & \times \left(\frac{h_\mathrm{peb}/h_\mathrm{g}}{0.15} \right)^{-1} \left(\frac{\tau_\mathrm{acc}}{5\,\mathrm{Myr}} \right)^{1/7} \left(\frac{M_\mathrm{p}}{M_\mathrm{Jup}} \right)^{-3/14} \left(\frac{r_\mathrm{out}/r_0}{16} \right)^{1/2},\\
\end{split}	
\end{equation}
where the numerical estimate is valid in the flaring part of the CPD (i.e., at $r > r_\mathrm{tran}$). 
In the above expression, $h_\mathrm{peb}$ is the aspect ratio of the pebble disk.
For a given pebble Stokes number and dimensionless vertical diffusion coefficient  $\delta$ \citep[see, e.g.,][and references therein]{Jo+14}, the pebble aspect ratio is a fraction of the gaseous disk aspect ratio such that
 \citep{DMS95}
\begin{equation}
	\frac{h_\mathrm{peb}}{h_\mathrm{g}} = \sqrt{\frac{\delta}{\mathrm{St}+\delta}},
\end{equation}
and we have used $\mathrm{St}=5\times10^{-3}$ and $\delta=10^{-4}$ to estimate $h_\mathrm{peb}$ in eq.~(\ref{q_mig}).

The value of $q_\mathrm{mig}$ depends only weakly on the mass of the planet and is therefore of the same order of magnitude in the Jovian and Saturnian CPDs. 
It however depends linearly on the pebble flux through the disk and the mass of the CPD.
For our choice of parameters, the migration mass lies one to two orders of magnitude below the mass of the major satellites of Jupiter and Saturn ($q_\mathrm{s} \sim 10^{-4}$).
Fig.~\ref{CPD_migration} presents numerical integrations of the growth tracks of protosatellites accreting pebbles while migrating inward through type-I migration. At first, growth largely dominates the evolution of the objects, but when approaching the migration mass, the growth curves of the protosatellites become nearly horizontal and they grow only little until they reach the inner edge of the CPD.

It is difficult to imagine that the mass of the CPDs were orders of magnitude lower, as this would render the capture and ablation of large planetesimals inefficient.
Given that the migration timescale of full-grown satellites is of the order of $10^3$--$10^4\,$years only, the survival of the satellites against migration would require that both their formation and the dissipation of the CPDs happened on even shorter timescales.
This conclusion is not very satisfying since, on the one hand, the dissipation mechanism of a circum-planetary disk and the timescale on which it operates are yet to be constrained, and on the other hand, such short accretion timescales for the moons would imply strong accretional heating, which would be difficult to reconcile with the inferred internal state of Callisto and Titan \citep[e.g.,][]{BC08}\footnote{We note that although there exists models of a differentiated Titan \citep{CRL10,OS14}, they still require limited heating during accretion to allow for the leaching of $^{40}$K (whose radioactive decay is the main long term energy source in the interior of the satellite) and the formation of hydrated minerals.}.

Instead, it seems that the migration of the protosatellites must have been stopped to allow for the formation and survival of the moons.
The most likely mechanism that could have prevented the protosatellites to migrate all the way down to the surface of their parent planet is the truncation of the CPD by an inner magnetic cavity \citep{SSI10,OI12}.
The sharp drop of density that would occur at the truncation of the disk is indeed known to act as a type-I migration trap \citep[e.g.,][]{MMCF06,LOL17}, thereby allowing planets or satellites to pile-up in resonant chains at the inner edge of the disk \citep[e.g.,][]{ODI10}.
Moreover, the magnetic coupling of the giant planets with their CPD and the opening of a magnetic cavity are the most promising mechanisms to explain their sub-critical rotation rate \citep{TS96,Ba18}.
Such a scenario would also place the formation and early dynamical evolution of giant planets' moons in the same context as that of compact systems around stars \citep[e.g.,][]{LOL17,Iz+17,Iz+19,La+19}.
We discuss that possibility below.

\subsection{Growth of satellites in resonant chains}

Protosatellites would rapidly migrate in the dense environment of the CPD and reach the inner edge of the disk before being able to grow to masses comparable to that of the observed satellite systems or the pebble isolation mass (Eq.~\ref{q_mig}, Fig.~\ref{CPD_migration}).
If the disk inner edge is truncated by a magnetosphere, a first migrating protosatellite would be trapped there, whereas a subsequently migrating object would be caught in a mean motion resonance with the inner protosatellite, resulting in the build-up of a resonant train of moons over time.

The protosatellites caught in such resonant chains would continue to grow until reaching the pebble isolation mass 
or up to the point when the giant planet has cleaned up its feeding zone and the influx of pebbles to the CPD stops.
In any case, most of the mass of the moons would be accreted while in the resonant chain, at a fixed distance from the planet. 

Given these considerations, we show in Fig.~\ref{Growth_map} growth maps of protosatellites at fixed radial distances in the Jovian and Saturnian CPDs, where we have assumed that pebbles have fixed Stokes numbers, with $\mathrm{St} = 5\times10^{-3}$, that the characteristic deposition radius of ablated material is $r_0 = 20\, R_\mathrm{Jup/Sat}$, and that $\dot{M}_0=3\times10^{-9}\,M_\mathrm{p}\, \mathrm{yr}^{-1}$. 
In every panel, the solid white line shows the time at which the pebble isolation mass is reached.

\begin{figure*}
	\centering
 \includegraphics[width=0.8\linewidth]{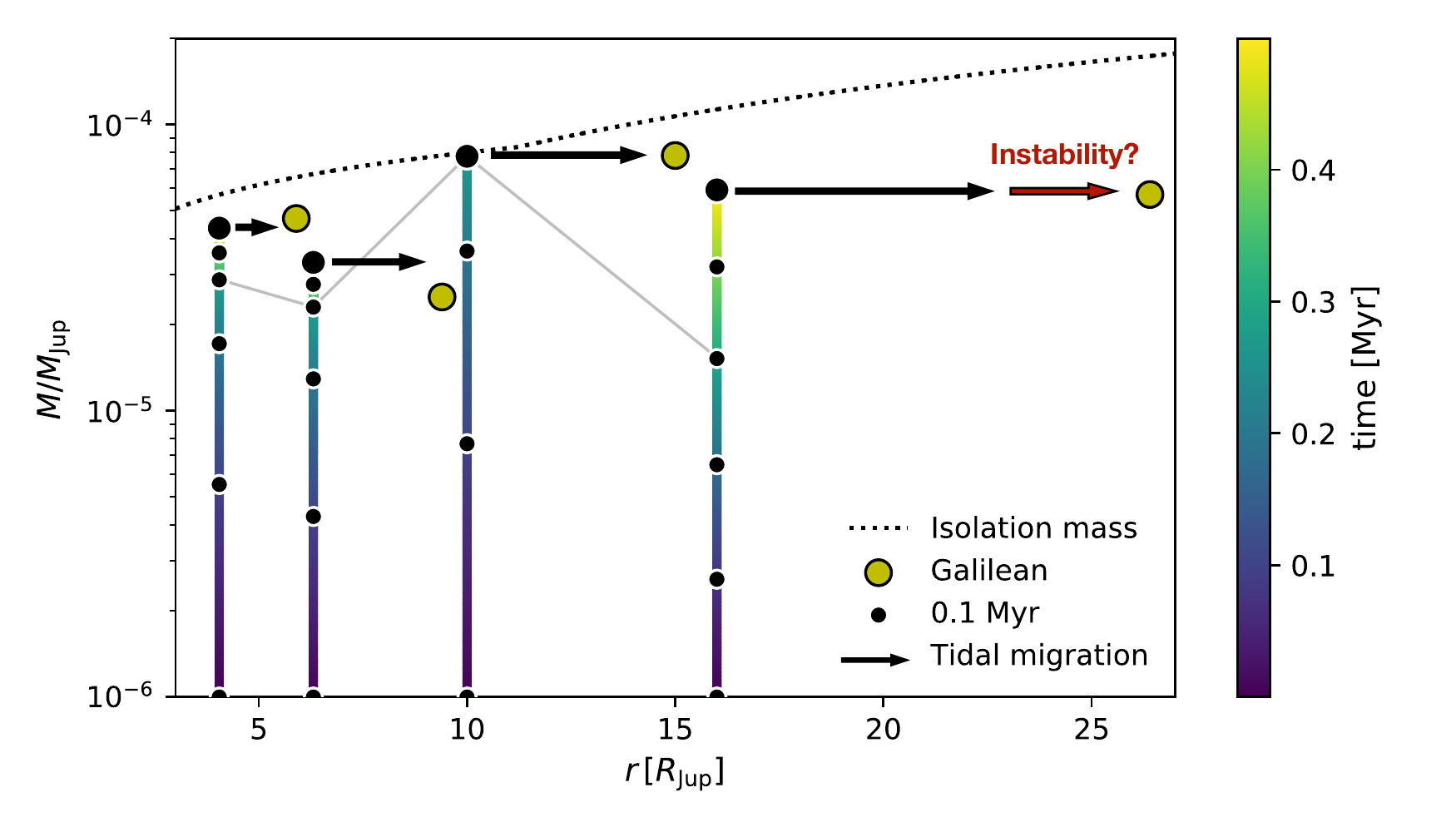}
 \caption{Example of pebble accretion growth tracks of Galilean moons analogues trapped in their mutual 2:1 mean motion resonances in a Jovian CPD with an accretion timescale onto Jupiter $\tau_\mathrm{acc}=10\,$Myr (see text for details). The growth tracks are shown as thick colored lines, where the color gives an indication of time. For further readability, black dots on the growth tracks show time intervals of 0.1 Myr and the larger dots indicate the end state of the system after 0.5 Myr of accretion. Ganymede here accretes at $10\,R_\mathrm{Jup}$, such that it is located beyond the snowline and the pebble isolation mass corresponds to the current mass of this moon. The thin gray line connects the system at the time Ganymede reaches the pebble isolation mass, which is after $\approx$0.3 Myr for our adopted value of $\dot{M}_0=3\times 10^{-9}\,M_\mathrm{p}\,\mathrm{yr}^{-1}$. After this point, the flux of pebbles inside of the orbit of Ganymede is greatly reduced, thereby slowing down the accretion of the inner moons, as indicated by the spacing between the black dots on their growth track being shortened. Here, the mass of the Galilean satellites (the current system is marked as yellow dots) is well reproduced, although the moons would have formed closer to Jupiter than where they are now. Tidal migration after the dissipation of the CPD can bring the orbits of the three inner satellites to their current location \citep{YP81}. Another mechanism, such as a dynamical instability, is however required to account for the fact that the outermost moon, Callisto, is presently outside of the resonance system.} \label{Growth_tracks}
\end{figure*}

The top panels of Fig.~\ref{Growth_map} are for cases in which the flux of pebbles through the CPD is simply given by Eq.~(\ref{Mdot_peb}).
Due to the smaller aspect ratio in the inner regions of the CPDs, pebble accretion is more efficient closer to the planet (see also Fig.~\ref{eff_PA_irr_visc}) and protosatellites grow faster there.
At distances beyond $r_0$, growth becomes very slow due to the exponential cut-off in the pebble flux and the larger aspect ratio of the CPDs.
The formation of compact satellite systems should thus be a quite natural outcome.

In the bottom panels of Fig.~\ref{Growth_map}, we have considered the effect of the snowline on the growth of the protosatellites, as this latter could have played an important role \citep[e.g.,][]{RMV17}.
We estimated that in the CPDs, the snowline lies at a temperature of $\sim$220--225$\,$K\footnote{This is assuming a water mass fraction of 0.571\%, corresponding to the solar system composition \citep{Lo03}. Here we define the position of the snowline as the distance where the equilibrium vapor pressure of water at temperature $T_\mathrm{d}$ matches the partial pressure of water in the CPD. If the CPDs are depleted (enriched) in water, the snowlines would lie at lower (higher) temperatures.}, corresponding to distances of $\sim$8$\,R_\mathrm{Jup}$ and $\sim$4$\,R_\mathrm{Sat}$ from Jupiter and Saturn, respectively.
We have not modeled the detailed processes, such as sublimation, condensation and diffusive redistribution, occuring at the snowline, but simply assumed that the flux of pebbles was halved inside of the snowline due to water evaporating from pebbles.
Including this reduction of the pebble flux, the region just outside the snowline becomes a sweet spot for the formation of moons, where protosatellites grow the fastest.
This could have some important implications since a protosatellite reaching the pebble isolation mass would substantially reduce the flux of pebbles inside of its orbit, thus hindering the growth of inner moons.
This effect is probably less so relevant in the case of cold CPDs, such as Saturn's, as the snowline stands very close to the planet. 

Applying such a scenario to the formation of the Galilean moons, we constructed pebble accretion growth tracks of Galilean analogues trapped in a chain of 2:1 mean motion resonances, including the reduction of the pebble flux due to outer moons catching some fraction of the pebbles.
These are presented in Fig.~\ref{Growth_tracks}.
Here we have assumed that Ganymede analogue is growing at $10\,R_\mathrm{jup}$ and that the aspect ratio of the CPD at this location is $h_\mathrm{g}=0.055$ (corresponding to assuming that $\tau_\mathrm{acc}=10\,$Myr according to Eq.~\ref{h_thin}), such that the pebble isolation mass equals the mass of Ganymede, and the snowline is located in between the orbits of Ganymede's analogue and Europa's analogue.
We note that the formation of the moons closer to Jupiter than where they are currently found is compatible with the subsequent tidal expansion of their orbits, at least in the case of the three inner ones that are part of the Laplace resonance system \citep{YP81, LAKH09}.
Another mechanism, such as a dynamical instability (see Sect.~\ref{Dyn_Insta}), is nevertheless necessary to explain that Callisto is currently not in resonance with the other moons.
Finally, we have assumed that $r_0 = 12.5\,R_\mathrm{Jup}$, that pebbles had a constant Stokes number $\mathrm{St}=5\times10^{-3}$, and stopped the integration when Callisto's analogue (the outermost moon) had reached the current mass of Callisto.

Importantly, the flux of pebbles inside of a moon having reached pebble isolation is not entirely cut-off because of material being constantly deposited by the ablation of planetesimals.
The flux of pebbles at a distance $r_1$ interior to a protosatellite that has reached pebble isolation at a distance $r_2$ is then
\begin{equation}
	\dot{M}_\mathrm{peb} = \dot{M}_0 \left( e^{-r_1^2/(2r^2_0)} - e^{-r_2^2/(2r^2_0)} \right).
\end{equation}
This still corresponds to an abrupt reduction of the pebble flux, as can be observed in the growth tracks of Io's and Europa's analogues in Fig.~\ref{Growth_tracks}, whose growth rates are substantially reduced after Ganymede's analogue has reached the pebble isolation mass.
At the end of the integration, the mass of Europa's analogue is $\approx$30\% higher than that of the actual satellite, and the mass of Io's analogue is only $\approx$8\% lower than the current mass of Io.

It is nonetheless important to note that the growth tracks presented here are merely illustrative and meant to demonstrate that a pebble accretion scenario, similar in many regards to the scenario recently proposed for the formation of close-in super-Earths \citep[e.g.,][]{La+19,Iz+19}, \textit{could} lead to the formation of a system similar to that of the Galilean moons. 
As discussed in Sect.~\ref{Dyn_Insta} however, a wide diversity of possible systems should be expected and their dynamical evolution must be explored with numerical simulations accounting for the mutual gravitational interactions among the satellites.

\begin{table*}
         \centering
     \caption[]{Mass and amount of pebbles required to grow Ganymede and Titan from an initial seed mass of $q_\mathrm{s}=10^{-8}$. The values in parentheses were computed for an aspect ratio $h_\mathrm{g}=0.1$, whereas the nominal values were calculated assuming that the pebble isolation mass corresponds to the mass of the satellite (eq.~\ref{q_iso}), that is $h_\mathrm{g}\approx0.055$ in the case of Ganymede, and $h_\mathrm{g}\approx0.08$ in the case of Titan.}
     \begin{tabular}{l c c c c}
    \hline \hline \vspace{2pt} 
     {}   &  $M_\mathrm{sat}/M_\mathrm{p}$ & \multicolumn{3}{c}{ $M_\mathrm{peb}/M_\mathrm{p}$} \vspace{2pt} \\
      \cline{3-5} \vspace{2pt} 
     \vspace{2pt} {} & {} & $\mathrm{St}=5\times10^{-3}$ & {} & $\mathrm{St}=5\times10^{-2}$\\
     \hline \vspace{2pt} 
    \vspace{2pt}Ganymede & $7.8\times10^{-5}$ & $1.0(5.2)\times10^{-3}$ & {} & $1.0(2.7)\times10^{-3}$\\
     Titan & $2.4\times10^{-4}$ & $3.3(6.1)\times10^{-3}$ & {} & $3.0(4.3)\times10^{-3}$\\
     \hline
     \end{tabular}\label{M_peb}
   \end{table*}
 
\section{Discussion}\label{discussion}

\subsection{Additional effects that can influence the capture and ablation of planetesimals}

\paragraph*{Aspect ratio of the CPD} Here we have assumed an average aspect ratio of the CPD of 0.06, appropriate for a cold disk around a slowly accreting jovian planet ($\tau_\mathrm{acc}=5\,$Myr). If CPDs are able to develop during earlier phases of the formation of a giant planet, when gas is accreted at a higher rate, they could be much thicker, with an average aspect ratio of 0.3 \citep{DAP15,Sz17}, which would influence the efficiency of capture and ablation of planetesimals. \citet{FOTS13} show that the energy of planetesimals captured on bound orbits is mostly dissipated at their closest approach from the planet. It is therefore possible to define a distance $r_\mathrm{capt}$ from the planet at which a sufficient amount of energy $\Delta E$ to allow capture is lost. The energy dissipation due to gas drag during closest approach is $\Delta E \approx a_\mathrm{drag} r_\mathrm{CA}$ \citep{TO10,FOTS13}, where $r_\mathrm{CA}$ is the distance at closest approach. 
Considering the case of neglible inclination (i.e., coplanar case), using eq.~\ref{a_drag}, \ref{tstop}, \ref{rho_PPD} and \ref{Sigma_CPD}, and setting $r_\mathrm{capt}=r_\mathrm{CA}$, we find
\begin{equation}
	\Delta E \approx \frac{3}{8} \frac{C_D v^2_\mathrm{rel}}{\rho_\mathrm{pl} R_\mathrm{pl}} \frac{\Sigma_\mathrm{out} r^{3/2}_\mathrm{out}}{2\pi h_\mathrm{g}} r^{-3/2}_\mathrm{capt}.
\end{equation}
It follows that $r_\mathrm{capt}\propto h^{-2/3}_\mathrm{g}$ if $v_\mathrm{rel}$ does not depend on the distance to the planet \citep[the dispersion regime of][]{FOTS13}. On the other hand, if planetesimals approach the planet at the escape velocity at distance $r_\mathrm{CA}$ \citep[the shear regime of][]{FOTS13} , then $v_\mathrm{rel} \propto r^{-1/2}_\mathrm{CA}$ and $r_\mathrm{capt}\propto h^{-5/2}_\mathrm{g}$. 
Although the dependency on the aspect ratio of the CPD is not very strong, it is noteworthy that gas drag assisted capture of planetesimals is more difficult in puffed-up disks. Thus, even if CPDs are able to form early on during the accretion history of gas giant planets, the capture and ablation of planetesimals might be inefficient until their CPDs cool down enough.

\paragraph*{Inclination of planetesimals} We have considered here the evolution of planetesimals with initially small inclinations, hence mostly experiencing coplanar interactions with the CPD of the giant planet. In the extreme opposite case, planetesimals would cross the CPD almost vertically, as discussed in \citet{MEC10}. Vertical disk crossers would interact with the gas in the CPD over a distance $\approx2H_\mathrm{g}$ \citep{MEC10} and the energy dissipated would then become $\Delta E \approx 2 a_\mathrm{drag} H_\mathrm{g}$.
Thus, comparing the amount of energy dissipated through gas drag in a coplanar encounter ($\Delta E_\parallel$) at distance $r_\mathrm{CA}$, with that dissipated in the case of a vertical encounter ($\Delta E_\perp$) at the same distance, we have crudely that $\Delta E_\perp / \Delta E_\parallel \sim H_\mathrm{g}/r_\mathrm{CA} = h_\mathrm{g} \ll 1$, where we have assumed that $v_\mathrm{rel}$ is of the same order of magnitude in both the coplanar and vertical encounters. 
The capture and ablation of highly inclined planetesimals should therefore be much less efficient than that of coplanar objects, which is consistent with the results of \citet{FOTS13}.
However, it was found in \citet{Ro18} that planetesimals could be delivered to the feeding zone of a giant planet with rather small inclinations ($i \lesssim 10^{-1}$) and effectively captured within the CPD.

\paragraph*{Break-up of planetesimals} The high dynamic (or ram) pressures experienced by planetesimals crossing the CPD at high relative speeds might cause them to break-up, an effect that has been ignored in the present study due to its complex treatment \citep[see, e.g.,][]{RMW17}. Fragmented planetesimals would have shorter ablation timescales which could change the distribution of material within the CPD. Assuming planetesimals are held together by their own gravity, for a given dynamic pressure $P_\mathrm{dy}=(1/2) \rho_\mathrm{g} v^2_\mathrm{rel}$ acting on them, they will be subject to break-up if their radius is smaller than \citep{PBT79}
\begin{equation}
	R_\mathrm{break-up} = \sqrt{\frac{5}{4\pi}\frac{P_\mathrm{dy}}{G \rho^2_\mathrm{pl}}}.
\end{equation}
It is then possible to derive the distance $r_\mathrm{break-up}$ at which planetesimals of a given size might break-up. In the case of a coplanar encounter, with $v_\mathrm{rel}=(\sqrt{2}\pm 1)v_\mathrm{K}$, and our jovian CPD with a constant aspect ratio $h_\mathrm{g}=0.06$, we find
\begin{equation}
r_\mathrm{break-up} \approx 4.3 \left( \sqrt{2} \pm 1 \right)^{4/7} \left(\frac{R_\mathrm{pl}}{100\,\mathrm{km}} \right)^{-4/7} R_\mathrm{Jup}.
\end{equation}
Therefore, fracturing and break-up of planetesimals is likely to occur within the CPDs of jovian planets, especially for the smaller planetesimals on retrograde orbits.
However, comparing with the curves showing the distribution of ablated material in Fig.~\ref{abl_distrib}, it appears that for every planetesimal size, break-up would occur in regions of the CPD where we find ablation to be an efficient mechanism, such that it is unlikely to significantly alter the distribution of dust in the disk. It should however be noted that break-up could play a more important role for the capture and ablation of highly inclined planetesimals which could otherwise be inefficient as discussed in the previous paragraph.

\subsection{Required amount of pebbles to grow moons and implications}\label{peb_eff}

Irrespective of the proposed mechanism of solids delivery to a CPD, it seems that pebble accretion is the most likely channel of growth for the moons. 
In the scenario proposed by \citet{CW02,CW06}, it was assumed that the small dust grains brought with the inflow of gas onto the CPD could grow up to large ($\gtrsim$km) sizes and avoid a rapid inward radial drift, but this should in fact not be the case \citep[e.g.,][and our Sec.~\ref{Dust_evol}]{SOSI17}.
Even if a dust trap would exist in CPDs, as recently proposed by \citet{DS18}, this would only allow for the formation of the satellite seeds which should subsequently grow by pebble accretion.
On the other hand, as investigated in Sec.~\ref{Capt_abl} \citep[see also][]{Es+09,MEC10,FOTS13}, the capture of planetesimals should mainly result in the delivery of small dust grains due to the strong ablation of the planetesimals crossing a CPD.
Finally, large icy building blocks would be incompatible with the observed composition gradient among the Galilean moons \citep{RMV17,DNOI13}, pointing towards the fact that pebbles would be better building blocks. 

In a pebble accretion scenario, however, the growing moons would only accrete a fraction of the drifting pebbles. 
A majority of pebbles would simply drift past the orbit of the moons, until one or more moons reach the pebble isolation mass and block the pebble flow outside their partial gas gap.
As shown by Fig.~\ref{eff_PA_irr_visc}, satellites comparable in mass to the Galilean moons or Titan catch $\approx$25\% of the pebble flux they encounter, and, for the integration shown in Fig.~\ref{Growth_tracks}, $1.5\times10^{-3}\,M_\mathrm{Jup}$ worth of pebbles were processed through the CPD to form a Galilean-like system whereas the actual mass of the system is about $2\times 10^{-4}\,M_\mathrm{Jup}$.

The total amount of pebbles necessary to grow a satellite can be expressed as
\begin{equation}
	f_\mathrm{peb} \equiv \frac{M_\mathrm{peb}}{M_\mathrm{p}} = \int^{q_\mathrm{sat}}_{q_\mathrm{seed}} \varepsilon^{-1}_\mathrm{PA} dq.
\end{equation}
Table~\ref{M_peb} reports the amount of pebbles needed to grow Ganymede and Titan, the most massive moons of Jupiter and Saturn, respectively.
Since smaller aspect ratios yield higher pebble accretion efficiencies, the minimum amount of pebbles required to grow a given satellite through pebble accretion is obtained when the aspect ratio of the disk is such that the pebble isolation mass corresponds to the mass of the satellite (a smaller aspect ratio would imply an isolation mass which is smaller than the mass of the satellite).
Even in this case, the overall (integrated) pebble accretion efficiency is less than 10\% for both Ganymede ($\approx$8\%) and Titan ($\approx$7\%), meaning that an amount of pebbles equivalent to more than 10 times the mass of the moons is required to allow for their formation (see Table~\ref{M_peb}).
Although this depends on the initial mass of the satellite seeds, considering seeds that are initially an order of magnitude more massive yields an increase in the integrated pebble accretion efficiency of only about 2\%.

It is therefore difficult to envision that many generations of satellites could form around giant planets such as proposed by \citet{CW06} \citep[and more recently by][]{Ci+18}, even if their CPD develops well before their completion.
If giant planets were able to efficiently cool down and contract while still accreting a lot of gas, allowing for a CPD  to form early on, this latter would likely be very puffed-up, with $h_\mathrm{g}\sim0.3$ \citep{DAP15,Sz17}.
Accretion would be extremely inefficient in such an environment, thus preventing the growth of satellites.

The low accretion efficiency of Galilean-like moons strengthens the argument that the inflow of material accreted by a giant planet towards the end of its formation can not provide enough solids to grow the observed satellite systems \citep{Ro18}. 
Dust grains larger than $\gtrsim$0.1$\,$mm are efficiently filtered by the gap opened by a giant planet in its native disk \citep{PM06,Paa07,ZNDEH12,WBGKP18}.
For typical dust size distributions, this would mean that a fraction of only 1\% to 10\% of the dust mass could be efficiently entrained with the gas accreted by the giant planet \citep{ZNDEH12}.
The formation of the satellites through the inflow of material accreted by the giant planet requires that a total mass  $M_\mathrm{tot}=f_\mathrm{peb}/\varepsilon_\mathrm{d}$ is processed through the CPD, where $\varepsilon_\mathrm{d}$ is the dust-to-gas mass ratio of the material accreted by the giant planet at the epoch of formation of the satellites.
If $\varepsilon_\mathrm{d}\sim10^{-3}$ (that is 10\% of the typical solar value), $M_\mathrm{tot}$ becomes comparable to or even larger than the mass of the planet.

\subsection{The putative role of dynamical instabilities}\label{Dyn_Insta}

In the present study, we have argued that the delivery of solids to a CPD through the capture and ablation of planetesimals yields to a formation scenario of moons which resembles closely that recently proposed for the formation of super-Earths \citep{Iz+19,La+19} and other compact systems around M-dwarf stars, such as the planets orbiting around Trappist-1 \citep{OLS17}.
Drawing further the analogy with super-Earth systems \citep[we note that in terms of mass ratio, the distribution of super-Earths peaks at a few $10^{-5}$, which is fairly similar to the massive moons of the Solar System;][]{Wu19, LLJ19}, the resonant chains formed through the pile-up of migrating protosatellites would be prone to dynamical instabilities \citep{Iz+17,Iz+19,La+19}.
These latter can result in mergers and breaking of orbital resonances.

The contrast between Saturn's satellite system, whose mass is largely concentrated in the massive moon Titan, and the Galilean system, composed by four roughly equal mass satellites, might be indicative of a more violent dynamical history around Saturn, where an initially, somewhat Galilean-like system, could have undergone an instability to finally merge into a single Titan while spawning mid-sized moons \citep{AR13,SG12}.

Even in the case of the Galilean system, it seems difficult to avoid that the outermost satellite Callisto was once part of the resonant chain of moons. 
A dynamical re-arrangement of the system after the dissipation of the CPD could have occurred as well. 

It thus appears important to further explore the formation of moons around giant planets through numerical simulations that can account for the gravitational interactions among the protosatellites and follow the long-term evolution of the formed systems (that is, after gas dissipation).
Here we have provided a parametrization that is relevant to this exploration and could be used to set up such numerical simulations.

\section{Summary}\label{sum}

In the past decade, new paradigms have emerged regarding several key processes in the theory of planet formation which challenge our current understanding of the formation of the giant planets' moons.
Firstly, the formation of large objects (i.e., planetesimals or satellitesimals) requires some special conditions to occur \citep[e.g.,][]{Jo+14}, which are not met in most parts of the circum-planetary disks where the moons should form \citep{SOSI17,DS18}. 
Secondly, as a giant planet carves a gap in its natal protoplanetary disk, it blocks the flux of inwardly drifting pebbles \citep{LJM14}, thereby isolating itself from the main source of solids needed to grow its satellites \citep{Ro18}. 
Lastly, the $\alpha$-disk model, commonly used to constrain the formation conditions of the moons \citep[e.g.,][]{CW02}, likely provides a poor description of the structure and evolution of a circum-planetary disk \citep{FOTI14,FKTG17,MBO19}.

Here we have thus considered the case of a low viscosity circum-planetary disk, where the main source of energy at the midplane comes from irradiation by the central planet.
As capture of planetesimals on initially heliocentric orbits should be the main solids delivery mechanism to the CPD in the late stages of formation of a giant planet \citep[e.g.,][]{Es+09,TOM12,TMM14,Ro18}, the fate of the captured objects was investigated considering their ablation through frictional heating.
We find that ablation is an efficient mechanism, resulting in the capture of large planetesimals in the CPD delivering primarily solids in the form of small dust grains.
A fraction of planetesimals can however survive ablation and remain as large objects ($\gtrsim \! 10\,$km) in the circum-planetary disk, from which protosatellites can start to form.

The dust grains provided by the continuous ablation of planetesimals grow rapidly into radially drifting pebbles, whose flux through the circum-planetary disk is regulated by the rate at which fresh material is supplied through ablation.
Protosatellites growth then mainly proceeds through the accretion of pebbles of millimeter-centimeter sizes.

We find that the overall pebble accretion efficiency ($\sim$10\%), and the fast migration timescale of full-grown moons in CPDs ($10^3$--$10^4\,$years), argue against a scenario where many generations of satellites are able to form around giant planets \citep{CW06,Ci+18}.
It seems more reasonable instead that the migration of protosatellites is stopped at the inner edge of the CPD due to, e.g., its truncation by a magnetic cavity \citep{TS96, SSI10, Ba18}. A similar conclusion was reached by \citet{Shib+19}.

It follows that giant planets' moons could have mainly grown while being caught in chains of orbital resonances, and their formation process would overall be very similar to that of super-Earths and other compact systems such as the Trappist-1 planets \citep{Iz+17,OLS17,La+19,Iz+19}.
The protosatellites would be able to grow until reaching the pebble isolation mass, which is comparable to the current masses of the Galilean satellites and Saturn's moon Titan (when the accretion timescale of the parent planet is $\gtrsim 5\,$Myr), or up to the point when the giant planet has cleaned its feeding zone and the influx of pebbles vanishes.

Given that resonant chains of planets are prone to dynamical instabilities \citep{Iz+17,Iz+19,La+19}, the formation and subsequent evolution of moon systems should be investigated using N-body simulations to quantify the importance of such dynamical events in setting a system's final architecture and, perhaps, understand the differences among the Jovian and Saturnian systems.

\begin{acknowledgements}
	The authors thank the anonymous referee for their careful reading of the manuscript and suggestions for improvement. TR is thankful to Michiel Lambrechts and Beibei Liu for helpful discussions. The authors also wish to warmly thank Alessandro Morbidelli for stimulating discussions that initiated this project. TR and AJ were supported by the European Research Council (ERC Consolidator Grant 724687-PLANETESYS). AJ was further supported by the Knut and Alice Wallenberg Foundation (grant  number 2012.0150) and the Swedish Research Council (grant  2018-04867).
\end{acknowledgements}

%
%

\bibliographystyle{aa}
\bibliography{draft}

\begin{appendix}
\section{Surface temperature of planetesimals}\label{App}

\begin{figure}
 \includegraphics[width=\linewidth]{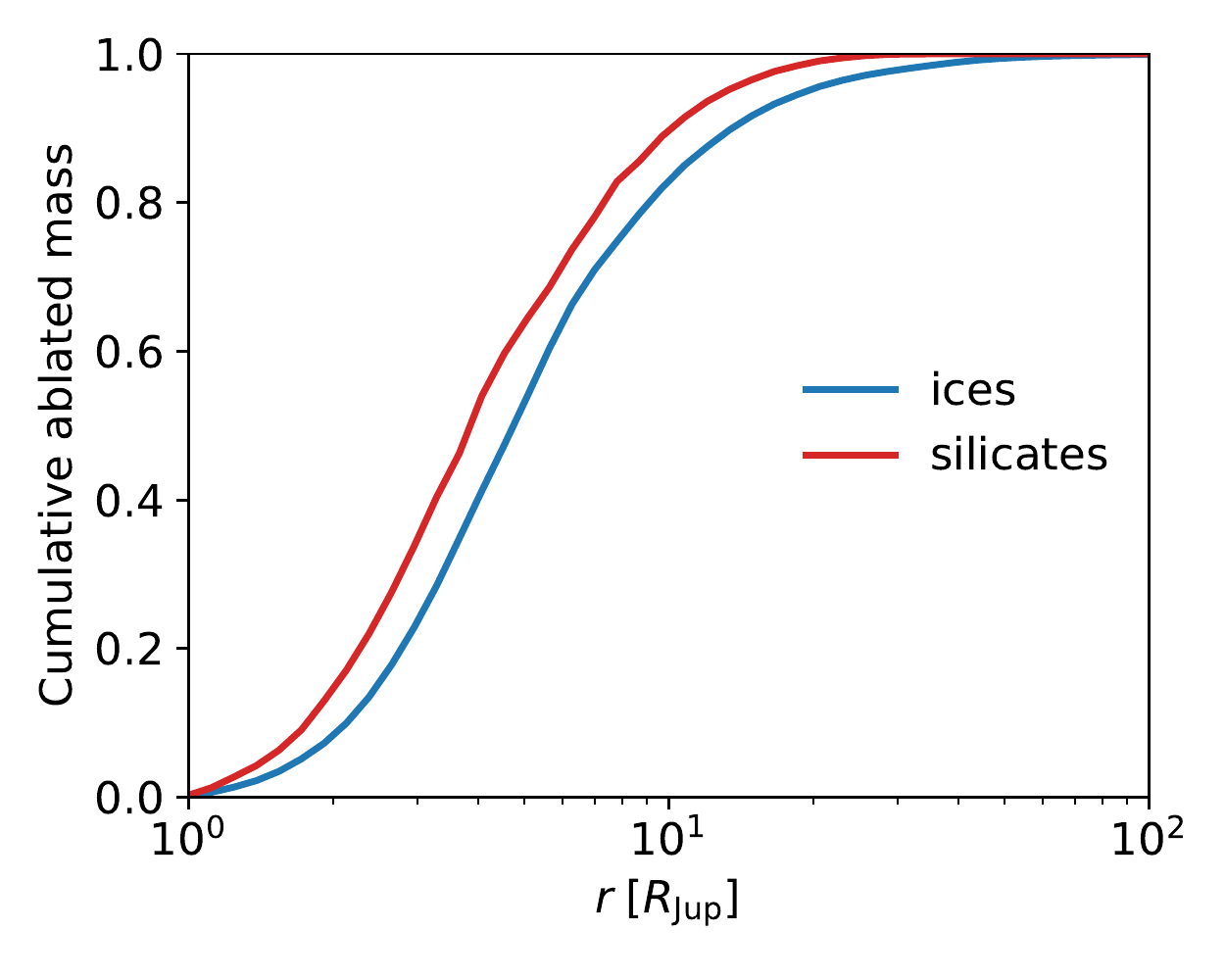}
 \caption{Comparison between the distribution of material ablated from the surface of $100\,$km sized planetesimals, considering ablation of water ice (blue curve), or silicates (red curve). For the silicates, we adopted the following parameters, $L_s = 5\times10^{6}\, \mathrm{J\,kg^{-1}}$, and $\log P_v =13.176-24605.0/T_\mathrm{pl}$ \citep[which is appropriate for enstatite;][]{Ca85}. The silicates are slightly more difficult to ablate than water ice, hence depositing material a bit closer to the planet. We note however that differentiated planetesimals would most likely be composed of hydrated minerals which are weaker than the silicates we have considered here \citep[e.g.,][]{Fe00}, which would bring the silicate curve closer to that of water ice. After a hundred planet's orbit, in the case of silicates, we find that the equivalent of $\sim$20\%  of the mass initially present in the feeding zone of the giant planet has been delivered in the CPD through ablation, compared to $\sim$25\% in the case of water ice.}\label{abl_ice_sil}
\end{figure}

The expression given in Eq.~(\ref{Tsurf}) for the surface temperature of planetesimals corresponds to the temperature at equilibrium between the different heating and cooling mechanisms, that is, when $dT_\mathrm{pl}/dt=0$, under given conditions (CPD temperature, density, and planetesimal's relative velocity with respect to the gas).
From \citet{DAP15}, the evolution of the surface temperature of a planetesimal evolves according to
\begin{equation}\label{Tsurf_full}
\begin{split}
\frac{4}{3}\pi \left[R^3_\mathrm{pl} - (R_\mathrm{pl} - \delta)^3 \right]\rho_\mathrm{pl} C_p \frac{dT_\mathrm{pl}}{dt} = & \, \frac{\pi}{8}C_D \rho_\mathrm{g}R^2_\mathrm{pl} v^3_\mathrm{rel} \\
{} & + 4\pi R^2_\mathrm{pl} \sigma_\mathrm{sb}(T^4_\mathrm{d}-T^4_\mathrm{pl}) \\
{} & + L_w \dot{m}_\mathrm{abl}\\
\end{split}
\end{equation}
where $C_p = 1600\, \mathrm{J\,kg^{-1}\,K}^{-1}$ is the heat capacity of the planetesimal, and 
\begin{equation}
\delta = 0.3 \frac{K_\mathrm{pl}}{\sigma_\mathrm{sb}T^3_\mathrm{pl}}
\end{equation}
is the thickness of the isothermal layer that is directly affected by the heating and cooling sources \citep{DAP15}, with $K_\mathrm{pl}= 3 \, \mathrm{W\,m^{-1}\,K^{-1}}$ is the thermal conductivity of the planetesimal.

\begin{figure}
 \includegraphics[width=\linewidth]{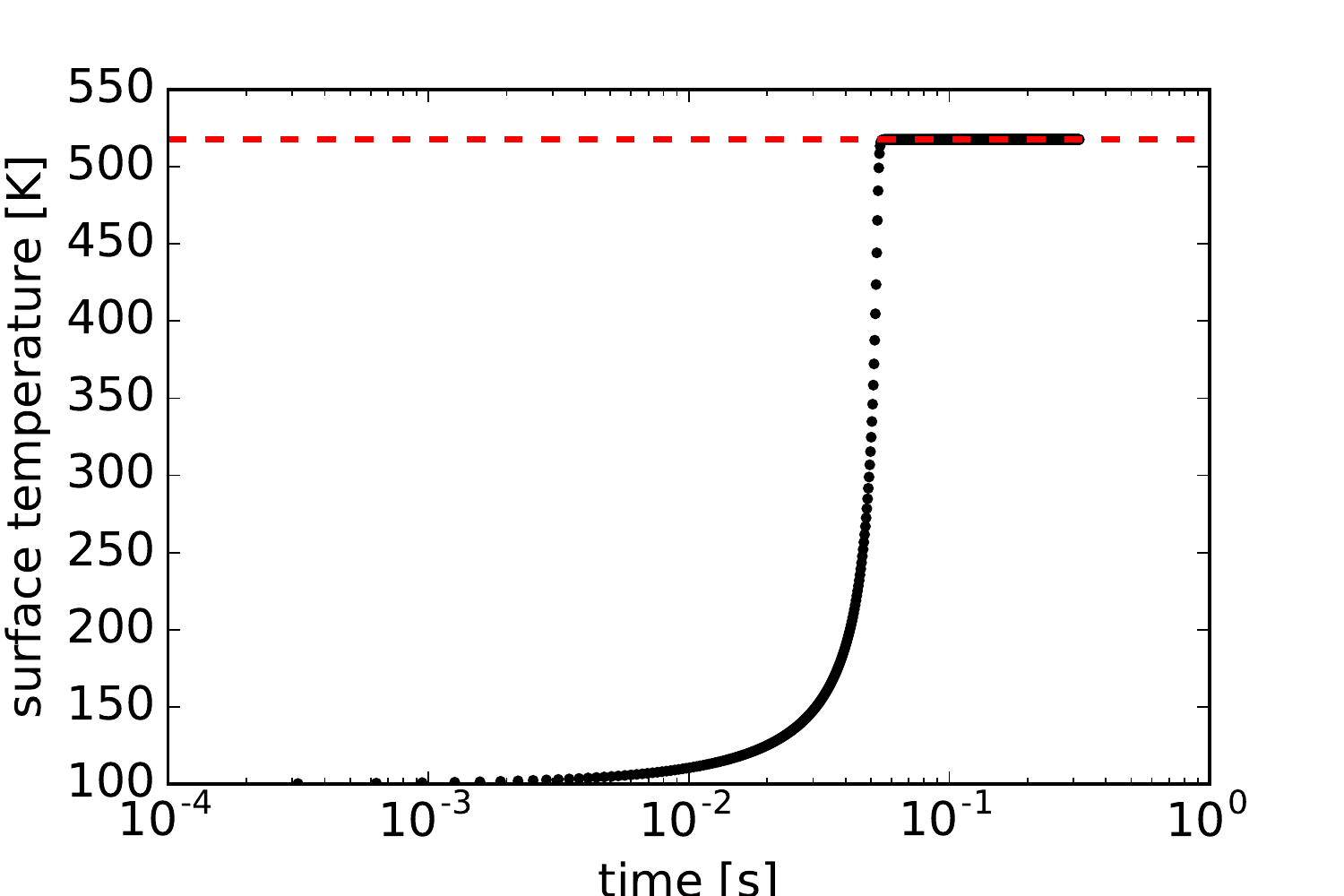}
 \caption{Comparison between the full integration of the evolution of the surface temperature of a planetesimal with a fourth order Runge-Kutta method (black dots), and the solution given by Eq.~(\ref{Tsurf}), shown as a red dashed line.}\label{T_surf_compar}
\end{figure}

In Fig.~\ref{T_surf_compar}, we show the results of the integration of Eq.~(\ref{Tsurf_full}) for a planetesimal with $R_\mathrm{pl}=100\,$km, using a 4th order Runge-Kutta method, starting from $T_\mathrm{pl}=100\,$K, with the parameters $T_\mathrm{d}=190\,$K, $\rho_\mathrm{g} = 4\times 10^{-2}\, \mathrm{kg\,m}^{-3}$, and $v_\mathrm{rel} = 33\, \mathrm{km\,s^{-1}}$, corresponding to a close approach at $10\,R_\mathrm{Jup}$ from Jupiter in the retrograde direction. 
We compare the results with the equilibrium solution given by Eq.~(\ref{Tsurf}) for the same parameters.
An equilibrium temperature is reached in $\sim$50$\,$ms only, which is much shorter than the typical perihelion passage timescale, of the order of a few hours (Eq.~\ref{t_CA}).
The solution given by Eq.~(\ref{Tsurf}) agrees perfectly with the equilibrium reached by solving Eq.~(\ref{Tsurf_full}), thus validating its use in our numerical integrations.

\section{Physical size of the dust particles in the CPD}\label{App_dust}
In the Jovian CPD, dust particles would transition from the Epstein to the Stokes drag regime at a size corresponding to \citep[e.g.,][]{We77}
\begin{equation}
s_\mathrm{Stokes} = \frac{9\lambda}{4} \approx 0.1\, \mathrm{mm}\, \left(\frac{r}{10\,R_\mathrm{Jup}} \right)^{3-\xi/2},
\end{equation}
where $\lambda \propto \rho_\mathrm{g}^{-1}$ is the mean free path of the gas molecules, and $\xi$ is the power-law index of the temperature in the CPD ($T_\mathrm{d}\propto r^{-\xi}$), which changes at the transition radius.
In this case, the size of the particles is given by
\begin{equation}
	s^\mathrm{(Stokes)}_\bullet = \sqrt{\frac{9\mathrm{St}h_\mathrm{g} r \lambda \rho_g}{4 \rho_\bullet}}.
\end{equation}
This is valid as long as the Reynolds number of the flow around the dust particles, $\mathrm{Re}=4s_\bullet v_\mathrm{rel}/c_\mathrm{g}\lambda$, lies below unity.
The further transition to the non-linear regime, at $\mathrm{Re}>1$, can only be approximately derived by analytical means, as $v_\mathrm{rel}$ is not known a priori.
However, since the Stokes number of the dust particles should remain small in the CPDs, their relative speed with respect to the gas in the azimuthal direction should vanish \citep[see, e.g., eq.~10 of ][]{Jo+14}, and $v_\mathrm{rel}$ should then be equivalent to the radial drift velocity of the particles, hence $v_\mathrm{rel}\approx 2 \mathrm{St} \eta v_\mathrm{K}$. 
The transition from the Stokes to the non-linear drag regime would then occur at a size corresponding to
\begin{equation}
s_\mathrm{NL} \approx \frac{c_\mathrm{g} \lambda}{8\mathrm{St}\eta v_\mathrm{K}} \approx 5.6\, \mathrm{mm}\, \left(\frac{10^{-2}}{\mathrm{St}} \right)  \left(\frac{r}{10\,R_\mathrm{Jup}} \right)^{3}.
\end{equation}
In this non-linear regime, the size of the particles is given by
\begin{equation}
 s^\mathrm{(NL)}_\bullet \approx \frac{9^{5/8}}{2^{1/2}} \frac{\mathrm{St}^{7/8} c^{3/8}_\mathrm{g} (\eta v_\mathrm{K})^{1/4}}{(\rho_\bullet \Omega_\mathrm{K})^{5/8} \rho^{1/8}_\mathrm{g}}.
\end{equation}

\section{Pebble accretion efficiency}\label{App3}

In the present study, we have used the pebble accretion efficiency as provided in \citet{LO18} and \citet{OL18}. 
It is expressed as a combination of the 2D and 3D regimes of pebble accretion,
\begin{equation}
\varepsilon_\mathrm{PA} = \left(\varepsilon^{-2}_\mathrm{2D} + \varepsilon^{-2}_\mathrm{3D} \right)^{-1/2}.
\end{equation}
In the above expression, $\varepsilon_\mathrm{2D}$ is the pebble accretion efficiency in the planar approximation (which is valid when the accretion radius is larger than the scale height of the pebbles), and reads \citep{LO18}
\begin{equation}
\varepsilon_\mathrm{2D} = 0.32 \sqrt{\frac{q_\mathrm{s}}{\eta^2 \mathrm{St}} \left(\frac{\Delta v}{v_\mathrm{K}} \right)},
\end{equation}
with $q_\mathrm{s}$ the ratio of the protosatellite's mass to that of the central planet, and $\Delta v$ the relative velocity between the accreting seed and the pebbles, which, following \citet{OL18}, can be expressed as
\begin{equation}
\Delta v = \left[1 + 5.7\left(\frac{q_\mathrm{s}\mathrm{St}}{\eta^3} \right) \right]^{-1} \eta v_\mathrm{K} + 0.52(q_\mathrm{s}\mathrm{St})^{1/3}v_\mathrm{K}.
\end{equation}
This expression ignores effects related to the eccentricity of the protosatellites, which is not considered here.
It also ignores the contribution of the gas accretion flow within the disk to the velocity of pebbles, which can become important for pebbles with small Stokes numbers \citep{JIB19}. However, expressing the velocity of the inward accretion flow as $v_\mathrm{acc}=M_\mathrm{p}/(2\pi r \Sigma_\mathrm{g} \tau_\mathrm{acc})$, we find it to be about an order of magnitude slower than the radial drift speed of pebbles with $\mathrm{St}=5\times10^{-3}$.

In the 3D regime, the pebble accretion efficiency is expressed as \citep{OL18},
\begin{equation}
\varepsilon_\mathrm{3D}= 0.39\frac{q_\mathrm{s}}{\eta h_\mathrm{peb}},
\end{equation}
where $h_\mathrm{peb}$ is the scale height of the pebbles, reading \citep{DMS95}
\begin{equation}
	\frac{h_\mathrm{peb}}{h_\mathrm{g}} = \sqrt{\frac{\delta}{\mathrm{St}+\delta}},
\end{equation}
where $\delta=10^{-4}$ is the vertical diffusion coefficient.
The small value of $\delta$ is motivated by the fact that the turbulence level in the CPD should be low due to the suppression of the magneto-rotational instability \citep{FKTG17}. 
Low turbulence levels are also expected in the case of protoplanetary disks \citep[e.g.,][]{Bai17}, which seems to be supported by observations measuring dust settling to levels of about 10\%  of the gas scale height \citep{Pi+16}.

\end{appendix}

\end{document}